\documentclass[12pt]{article}
\usepackage{graphicx}
\usepackage{amsmath}
\usepackage{amsthm}
\usepackage{amssymb}
\usepackage{subcaption}
\usepackage{amsfonts}
\usepackage{float}    
\usepackage{xcolor}
\usepackage{booktabs} 
\usepackage{dsfont}
\usepackage{hyperref}
\usepackage{comment}
\date{}

\usepackage{enumerate}
\usepackage{natbib} 
\usepackage{url} 
\usepackage{enumitem} 
\setlist[itemize]{noitemsep}
\setlist[enumerate]{noitemsep}
\newcommand{\blind}{0}

\newtheorem{definition}{Definition}[section]
\newtheorem{theorem}{Theorem}[section]

\newtheorem{proposition}{Proposition}[section]
\newtheorem{remark}{Remark}[section]

\date{}

\begin{document}

\def\spacingset#1{\renewcommand{\baselinestretch}%
{#1}\small\normalsize} \spacingset{1}


\if0\blind 
{
  \title{\bf Likelihood-Free Inference for Multivariate Generalized Pareto Models}
  \author{Samira Aka\hspace{.2cm}\\
    LSCE, 
 Universit\'e Paris Saclay \& ESSEC Business School CREAR\\
    and \\
    Marie Kratz \\
    ESSEC Business School, IDO Dep. and CREAR\\
    and \\
    Philippe Naveau 
    \\ Laboratoire\,des\,Sciences\,du\,Climat\,et\,de\,l’Environnement, CNRS-CEA-UVSQ-IPSL}
  \maketitle
} \fi

\if1\blind
{
  \bigskip
  \bigskip
  \bigskip
  \begin{center}
    {\LARGE\bf Title}
\end{center}
  \medskip
} \fi

\bigskip
\begin{abstract}
\noindent Likelihood-based inference for multivariate extreme-value models is often unreliable or infeasible when likelihoods are intractable or supports are discrete. This challenge is particularly acute for multivariate discrete generalized Pareto models, where both marginal tail behavior and dependence must be inferred from sparse exceedance samples. We propose a two-stage likelihood-free inference procedure, termed \emph{AW--NBE} 
(Adaptive Wasserstein Neural Bayes Estimator), that combines neural Bayes estimation 
with a targeted optimal transport refinement step based on the Sinkhorn discrepancy. In the first stage, a neural Bayes estimator trained on simulated data provides fast and stable initial parameter estimates.  In the second stage, these estimates are locally refined by minimizing the Sinkhorn 
divergence between the empirical distributions of observed and simulated exceedances. This refinement reduces the Sinkhorn discrepancy between the empirical distributions of observed and simulated exceedances, while preserving dependence features learned by the neural estimator.
Model adequacy is assessed using new optimal transport based multivariate Q--Q and potential diagnostics.
Applications to financial log-returns and Swiss dry spell exceedances suggest that AW--NBE can improve parameter inferences compared to estimation using solely, either the Sinkhorn discrepancy, or the standard neural Bayes estimators and censored likelihood estimation.
\end{abstract}

\vfill
\newpage
\section{Introduction}

Statistical inference for parametric models commonly relies on likelihood-based methods such as maximum likelihood estimation (MLE) or composite likelihood \citep{lindsay1988,cox2004theory,varin2011}, which are widely used due to their theoretical properties and efficiency. 
These approaches face significant limitations when the likelihood is intractable or computationally expensive to evaluate, as is often the case for multivariate models with complex dependence structures or discrete support. 
In such settings, MLE may be infeasible and composite likelihood methods are commonly used as practical surrogates.

A large class of likelihood-free techniques instead relies on simulation-based comparisons between observed data and synthetic data generated under candidate parameter values \citep{Papamakarios2019,Cranmer2020}. 
Among these, approximate Bayesian computation (ABC) replaces likelihood evaluation by accepting parameters that generate simulated data sufficiently close to observed data, typically measured through summary statistics \citep{Beaumont2002, Lintusaari2017}. 
Synthetic likelihood methods approximate the likelihood of summary statistics through a parametric model, while indirect inference relies on auxiliary models to link simulated and observed data \citep{Gourieroux1993, Heggland2004, Wood2010, Price2018}.

More recently, deep learning simulation-based inference methods have been proposed, enabling inference through neural networks trained on simulated data \citep{sainsbury2024neural}. 
These include neural posterior and likelihood approximation methods \citep{Rezende15,Papamakarios2021}, as well as neural Bayes estimators (NBE) \cite[see, e.g.][]{sainsbury2024neural}, which directly approximate Bayes estimators under a given loss function. 
NBE provide a flexible and computationally efficient framework, particularly well suited to complex dependence structures and high-dimensional observations, while allowing fast evaluation once trained.

An alternative class of likelihood-free methods relies on distributional distances between probability measures. 
Optimal transport (OT) provides a principled framework to compare probability distributions through Wasserstein-type discrepancies \citep{villani2008optimal, Panaretos2019, Bernton2019}. 
Such approaches capture global distributional features and can naturally accommodate dependence and heavy tails. 
To address computational challenges, scalable approximations such as entropy-regularized optimal transport (EOT), also known as Sinkhorn divergence-based optimal transport, have been proposed \citep{cuturi2013sinkhorn, Genevay2018, flamary2021pot}, together with projection-based variants such as sliced Wasserstein distances \citep{Bonneel2015, Nietert2022}.
These approaches exhibit complementary strengths. NBEs are fast and scalable but may lack accuracy in the tail, while OT-based methods directly enforce distributional alignment but require solving a potentially challenging optimization problem. 
This complementarity is particularly relevant in extreme-value settings.

In multivariate extreme value theory (EVT) (see \cite{Resnick1987}, \cite{Resnick2007}, \cite{DeHaanFerreira2006}, Chap.~7 by Naveau and Segers in \cite{HandbookExtremes2026}), threshold exceedances are naturally modeled by multivariate generalized Pareto distributions (MGPDs) \citep{rootzen_tajvidi2006, rootzen_segers_wadsworth, rootzenSegerswadsworth_2, Kiriliouk2019}. 
Although MGPDs provide a theoretically sound framework for modeling joint tail behavior, their likelihood is often unavailable in closed form for general dependence structures, making inference challenging, especially in high dimensions or in discrete settings.
Moreover, in various applications, including climate science, environmental monitoring, and insurance, extreme events are naturally recorded as counts or durations, such as threshold exceedance counts, or damage reports. These discrete-valued extremes are not naturally accommodated within continuous MGPD models without additional modeling assumptions. To address this gap, discrete extensions of the MGPD have been recently proposed, such as the multivariate discrete generalized Pareto Distribution (MDGPD) in  \cite{aka2025multivariatediscretegeneralizedpareto}, designed specifically for modeling multivariate extremes with integer-valued supports. 
The MDGPD lacks a closed-form density and often relies on latent-variable or simulation-based methods. As a result, standard likelihood-based approaches are no longer applicable and alternative inference strategies are needed. The statistical challenges arising in these discrete constructions go beyond those encountered in continuous MGPD models. 
Discrete supports generate irregular likelihood structures and combinatorial features that complicate likelihood-based calibration \citep{Chakraborty2015,hitz_davis_samorodnitsky_2024}. 
Moreover, as is well-known in multivariate extremes, the rarity of joint exceedances further reduces the effective sample size after thresholding \citep{Coles2001,DeHaanFerreira2006}. 
These limitations naturally motivate the use of simulation-based and likelihood-free inference methodologies that rely on forward model simulation rather than explicit likelihood evaluations.\\[1ex]
In this work, we develop a likelihood-free inference strategy 
that combines neural Bayes estimation with a targeted optimal transport refinement that might be beneficial, especially, in the discrete setting. 
The proposed approach, denoted AW--NBE for adaptive Wasserstein-NBE, first provides fast parameter estimation through simulation-based neural training and then improves calibration by minimizing a Sinkhorn discrepancy between observed and simulated exceedances \cite[see e.g.,][]{Feydy2019,Genevay2018}. 
The framework is applicable to both continuous and discrete exceedance models and is supported by theoretical results and empirical investigations on financial and environmental datasets, with gains that are especially significant in the discrete setting.

The main contributions in this paper are the following:
\begin{itemize}
\item A new hybrid likelihood-free estimator, $\widehat{\boldsymbol{\theta}}_n^{\mathrm{AW}}$, combining neural Bayes inference and optimal transport refinement;
\item A finite-sample guarantee that AW--NBE does not increase the empirical Sinkhorn discrepancy compared to NBE;
\item Consistency and, under a local regularity condition, asymptotic normality of AW--NBE, with asymptotic equivalence to the EOT estimator;
\item Empirical evidence, on a continuous financial dataset and a discrete environmental dataset, that the proposed refinement is most beneficial in the discrete setting.
\end{itemize}

The paper is organized as follows: Section~\ref{sec:likfree} presents the likelihood-free methods compared in the paper,
specifically the novel AW--NBE estimator and its theoretical properties. 
Section~\ref{sec:real_data} illustrates the practical relevance of the approach through applications to real datasets. Proofs of the main theoretical results are deferred to the Appendix.
\\[1ex]
Before concluding this introduction, let us introduce some notations. Bold face symbols denote vectors in $\mathbb{K}^d$, where $\mathbb{K}=\mathbb{R},\mathbb{Z}$ or $\mathbb{N}$, for example, $\boldsymbol{0} = (0, \ldots, 0)$ and $\boldsymbol{1} = (1, \ldots, 1)$. Mathematical operations such as addition, multiplication, and exponentiation are to be interpreted componentwise when applied to vectors. For instance, for $\boldsymbol{x}, \boldsymbol{\xi} \in \mathbb{R}^d$, we write $\left(\boldsymbol{1}+\boldsymbol{\xi}\boldsymbol{x}\right)^{-\boldsymbol{1}/\boldsymbol{\xi}}$ for the vector 
$\left((1 + \xi_1 x_1)^{-\frac{1}{\xi_1}}, \ldots, (1 +  \xi_d x_d)^{-\frac{1}{\xi_d}}\right)$.

We denote by $P_{\boldsymbol{\theta}}$ the distribution of the observations under parameter $\boldsymbol{\theta}$, and by $\pi(\boldsymbol{\theta})$ the prior distribution.  We let $\varepsilon>0$ denote the entropic regularization parameter.
 Writing $\left(\boldsymbol{Z}\nleqslant \boldsymbol{u}\right)$ means that at least one coordinate of $\boldsymbol{Z}$ exceeds the corresponding coordinate of $\boldsymbol{u}$, i.e., $\bigvee_{j=1}^d \left( Z_j > u_j \right)
$. $\mathds{1}_{A}$ is the indicator function, equal to 1 if $A$ is true and 0 otherwise. Given random elements $(\boldsymbol{\theta}, \boldsymbol{X})$, we denote by $\mathbb{E}[\boldsymbol{\theta} \mid \boldsymbol{X}]$ the conditional expectation of $\boldsymbol{\theta}$ given $\boldsymbol{X}$. 
For any generic function $f_\psi$ parameterized by $\psi$, we denote by $f_{\widehat{\psi}}$ its calibrated version obtained from simulated data. Finally, estimators of the parameter $\boldsymbol{\theta}$ computed from a sample of size $n$ are denoted by $\widehat{\boldsymbol{\theta}}_n^{\mathrm{Method}}$, where the superscript indicates the name of the chosen estimation method.

\section{Likelihood-free inference}\label{sec:likfree}

Throughout this section, let $\{P_{\boldsymbol{\theta}} : \boldsymbol{\theta} \in \Theta 
\subset \mathbb{R}^p\}$ be a parametric statistical model on $\mathbb{R}^d$, and let 
$P_{\boldsymbol{\theta}_0}$ denote the true distribution. We observe 
$\boldsymbol{X}_1,\dots,\boldsymbol{X}_n \overset{i.i.d.}{\sim} P_{\boldsymbol{\theta}_0}$, 
with empirical measure $\widehat{P}_n = \frac{1}{n}\sum_{i=1}^n \delta_{\boldsymbol{X}_i}$. 
For each $\boldsymbol{\theta} \in \Theta$, we assume that one can simulate 
$\boldsymbol{Y}^{(\boldsymbol{\theta})}_{1},\dots,\boldsymbol{Y}^{(\boldsymbol{\theta})}_{m_n}
\overset{i.i.d.}{\sim} P_{\boldsymbol{\theta}}$, with empirical measure 
$\widehat{P}_{m_n, \boldsymbol{\theta}} = \frac{1}{m_n}\sum_{j=1}^{m_n} 
\delta_{\boldsymbol{Y}^{(\boldsymbol{\theta})}_j}$; this assumption is standard in 
likelihood-free inference \citep{Cranmer2020, Papamakarios2019}.

\subsection{Background: neural Bayes and optimal transport approaches}

Likelihood-free inference has recently benefited from simulation-based methods that compare observed data with synthetic samples generated under candidate parameter values. 
A prominent approach is the NBE introduced by \citet{sainsbury2024neural}, which approximates the Bayes estimator using a neural network trained on simulated data. 
Let $\boldsymbol{\theta}\in\Theta$ denote the model parameter and 
$\boldsymbol{X}_{n,d}=(\boldsymbol{X}_1,\dots,\boldsymbol{X}_n)$ be an observed sample 
from $P_{\boldsymbol{\theta}}$, where each $\boldsymbol{X}_i \in \mathbb{R}^d$ 
and $n$ denotes the sample size.
Training data are generated by simulating pairs 
$(\boldsymbol{\theta}^{(k)}, \boldsymbol{X}_{n,d}^{(k)})$, where 
$\boldsymbol{\theta}^{(k)}\sim \pi(\boldsymbol{\theta})$ and 
$\boldsymbol{X}_{n,d}^{(k)}\sim P_{\boldsymbol{\theta}^{(k)}}$.
A neural network $f_\psi$, where $\psi$ denotes its parameter, is trained to minimize
\[
\frac{1}{K}\sum_{k=1}^K \left\| f_\psi(\boldsymbol{X}_{n,d}^{(k)}) - \boldsymbol{\theta}^{(k)} \right\|^2, \text{ where } K \in \mathbb{N} \setminus \{0\},
\]
so that, under quadratic loss, it approximates $\mathbb{E}[\boldsymbol{\theta}\mid \boldsymbol{X}]$. 
The resulting estimator is
\begin{equation}\label{eq:theta_NBE}
\widehat{\boldsymbol{\theta}}_n^{\mathrm{NBE}} = f_{\widehat{\psi}}(\boldsymbol{X}_{n,d}).
\end{equation}
This approach is computationally efficient once trained and well suited to complex or high-dimensional models.

An alternative class of likelihood-free methods relies on distributional discrepancies 
between empirical measures. In this work, we use the Sinkhorn divergence $S_\varepsilon$  
as a computationally tractable optimal transport criterion \citep{cuturi2013sinkhorn, Genevay2018, 
Feydy2019}; see Equation~\eqref{eq:Sinkhorn}. Given an observed sample $\boldsymbol{X}_{n,d}$
and simulated samples $\boldsymbol{Y}_1^{(\boldsymbol{\theta})},\dots,
\boldsymbol{Y}_m^{(\boldsymbol{\theta})}\sim P_{\boldsymbol{\theta}}$ 
with respective empirical measures 
\begin{equation}\label{eq:empirical_measures}
\widehat P_n = \frac1n \sum_{i=1}^n \delta_{\boldsymbol{X}_i}
\quad \text{and} \quad
\widehat P_{m,\boldsymbol{\theta}} =
\frac1{m} \sum_{j=1}^{m} \delta_{\boldsymbol{Y}^{(\boldsymbol{\theta})}_j},
\end{equation}
the transport-based estimator is defined as
\[
\widehat{\boldsymbol{\theta}}_n^{\mathrm{EOT}}
\in
\arg\min_{\boldsymbol{\theta}\in\Theta}
S_\varepsilon(\widehat P_n,\widehat P_{m,\boldsymbol{\theta}}),
\]
whenever a minimizer exists \cite[see, e.g.][]{cuturi2013sinkhorn,Genevay2018,Bernton2019}. 
Such estimators only require the ability to simulate from the model and can capture global distributional features, including dependence and tail behavior. In the definition of the EOT estimator, $m$ denotes a generic simulated sample size,
whereas in the asymptotic analysis we write $m_n$ to emphasize that the simulated
sample size may depend on $n$ and grows with $n$.

These two approaches exhibit complementary strengths: NBE provides fast and stable estimation but does not explicitly enforce distributional alignment, while optimal transport directly targets discrepancies between empirical distributions but may involve a more challenging optimization problem. 
One main question is how to merge these two complementary approaches. 

\subsection{Improving neural Bayes inference with optimal transport: Adaptive Wasserstein Neural Bayes Estimator}
\label{sec:awnbe}

Although our final statistical objective is to estimate MGPD parameters (i.e. modeling multivariate extremes), this section provides likelihood-free inference procedures  that can be applied beyond the MGPD class. 
 More precisely, we define an adaptive Wasserstein neural Bayes estimator (AW--NBE), which can be interpreted as a regularized optimal transport refinement of a NBE. 
Let $\Pi(\mu,\nu)$ denote the set of couplings between probability measures
$\mu$ and $\nu$ on $\mathbb{R}^d$ and $c:\mathbb{R}^d \times \mathbb{R}^d \to [0,+\infty)$
be a measurable cost function, typically chosen as $c(x,y)=\|x-y\|^p$ for some $p \geq 1$.
The entropy-regularized optimal transport cost between $\mu$ and $\nu$ is defined as
\[
\mathrm{OT}_\varepsilon(\mu,\nu)
=
\inf_{\pi \in \Pi(\mu,\nu)}
\left\{
\int c(x,y)\, d\pi(x,y)
+
\varepsilon\,\mathrm{KL}(\pi\,\|\,\mu\otimes\nu)
\right\},
\]
where $\mathrm{KL}$ denotes the Kullback–Leibler divergence and $\varepsilon>0$ the entropic regularization parameter.
We denote by $S_\varepsilon(\mu,\nu)$ the Sinkhorn divergence between probability measures $\mu$ and $\nu$
\cite[see, e.g.][]{Genevay2018,Feydy2019}, defined by: 
\begin{equation}\label{eq:Sinkhorn}
S_\varepsilon(\mu,\nu)
=
\mathrm{OT}_\varepsilon(\mu,\nu)
-
\frac12 \mathrm{OT}_\varepsilon(\mu,\mu)
-
\frac12 \mathrm{OT}_\varepsilon(\nu,\nu).
\end{equation}

This quantity defines a non-negative discrepancy between probability measures
that interpolates between optimal transport distances and kernel-based
integral probability metrics
\citep{Genevay2018,Feydy2019}. In particular,
\[
S_\varepsilon(\mu,\nu)\ge 0
\quad \text{and}\quad
S_\varepsilon(\mu,\mu)=0.
\]

In what follows, $S_\varepsilon(\cdot,\cdot)$ measures the discrepancy between
the empirical observed and model-simulated distributions.
\begin{definition}[Adaptive Wasserstein Neural Bayes estimator]
\label{def:AWNBE_sinkhorn}

For $\lambda_n \ge 0$, define
$$
Q_n(\boldsymbol{\theta})
=
S_\varepsilon(\widehat P_n,\widehat P_{m_n,\boldsymbol{\theta}})
+
\lambda_n \|\boldsymbol{\theta}-\widehat{\boldsymbol{\theta}}_n^{\mathrm{NBE}}\|_2^2,
$$
where $S_{\varepsilon}$ is defined in Equation~\eqref{eq:Sinkhorn},  $\widehat{\boldsymbol{\theta}}_n^{\mathrm{NBE}}$ in Equation~\eqref{eq:theta_NBE}, $\widehat P_n$, $\widehat P_{m_n,\boldsymbol{\theta}}$  in Equation~\eqref{eq:empirical_measures} and $\|\cdot\|_2$ is the Euclidean norm.
Any minimizer of $Q_n$, whenever it exists,
$$
\widehat{\boldsymbol{\theta}}_n^{\mathrm{AW}}
\in
\arg\min_{\boldsymbol{\theta}\in\Theta} Q_n(\boldsymbol{\theta})
$$
is called an adaptive Wasserstein NBE (AW--NBE).
\end{definition}

\begin{remark}
The use of the Euclidean norm $\|\cdot\|_2$ is mainly required for the local asymptotic analysis in Theorem~\ref{thm:asymptotics_AWNBE}, 
where second-order expansions and quadratic terms are involved. 
The other results remain valid for more general norms on $\mathbb{R}^p$.
\end{remark}
The following result establishes consistency  of AW--NBE. Throughout the asymptotic analysis, $(m_n)_{n \geq 1}$ denotes a sequence of positive integers with $m_n \to \infty$ as $n \to \infty$, representing the simulation sample size; its growth rate relative to $n$ is made precise in (H3)(i) and strengthened in $(H4^*)$. 
The result holds under the following hypotheses:
\begin{enumerate}
\item[(H1)] $\Theta\subset\mathbb{R}^p$ is compact.
\item[(H2)] (Identification)
\[
S_\varepsilon(P_{\boldsymbol{\theta}_0},P_{\boldsymbol{\theta}})>0,
\quad\forall\boldsymbol{\theta}\in\Theta\setminus\{\boldsymbol{\theta}_0\}.
\]
\item[(H3)] (Uniform convergence and continuity)
\begin{enumerate}
\item[(i)] (Uniform convergence) As $n\to\infty$ with $m_n\to\infty$,
\[
\sup_{\boldsymbol{\theta}\in\Theta}
\Big|
S_\varepsilon(\widehat P_n,\widehat P_{m_n,\boldsymbol{\theta}})
-
S_\varepsilon(P_{\boldsymbol{\theta}_0},P_{\boldsymbol{\theta}})
\Big|
\xrightarrow{\mathbb P}0.
\]
\item[(ii)]  (Continuity) The mapping
$\boldsymbol{\theta} \longmapsto S_\varepsilon(P_{\boldsymbol{\theta}_0},P_{\boldsymbol{\theta}})$
is continuous on $\Theta$.
\end{enumerate}
\item[(H4)] As $n\to\infty$, $\lambda_n \to 0$.
\end{enumerate}

These assumptions, standard in the analysis of extremum estimators \cite[see, e.g.,][]{vandervaart1998}, ensure
identifiability, stability of the objective function, and asymptotic vanishing of the regularization term.

\begin{theorem}\label{thm:consistency_aw_nbe}
Under (H1)--(H4), as $n\to \infty,$
\[
\widehat{\boldsymbol{\theta}}_n^{\mathrm{AW}}
\xrightarrow{\mathbb P}
\boldsymbol{\theta}_0.
\]
\end{theorem}
The proof of Theorem~\ref{thm:consistency_aw_nbe} relies on the standard separation argument for extremum estimators \cite[see, e.g.,][]{NeweyMcFadden1994, vandervaart1998} and is provided in Appendix~\ref{app:consistency_aw_nbe}.
After proving consistency, we now investigate the local asymptotic behavior of $\widehat{\boldsymbol{\theta}}_n^{\mathrm{AW}}$. 
Since a direct analysis is currently not feasible due to the lack of known convergence rates for the NBE, we adopt a comparison-based approach and relate $\widehat{\boldsymbol{\theta}}_n^{\mathrm{AW}}$ to our two benchmark estimators: the transport-based estimator $\widehat{\boldsymbol{\theta}}_n^{\mathrm{EOT}}$ and the NBE $\widehat{\boldsymbol{\theta}}_n^{\mathrm{NBE}}$.

The asymptotic theory of entropic optimal transport functionals is now well understood,
in particular through Hadamard differentiability and functional delta method arguments.
Combined with standard M-estimation theory (\cite{vandervaart1998}), it implies that, under suitable regularity,
smoothness, and identification conditions,
\begin{equation}\label{eq:EOT_asymptotics}
\sqrt n\bigl(\widehat{\boldsymbol{\theta}}_n^{\mathrm{EOT}} - \boldsymbol{\theta}_0\bigr)
\xrightarrow[n\to\infty]{d}
\mathcal N\!\left(
0,\,
\mathbf V_\varepsilon(\boldsymbol{\theta}_0)^{-1}
\mathbf\Sigma_\varepsilon(\boldsymbol{\theta}_0)
\mathbf V_\varepsilon(\boldsymbol{\theta}_0)^{-1}
\right),
\end{equation}
where $\mathbf V_\varepsilon(\boldsymbol{\theta}_0)$ denotes the Hessian of the limiting objective function 
$Q(\boldsymbol{\theta}) = S_\varepsilon(P_{\boldsymbol{\theta}_0}, P_{\boldsymbol{\theta}})$ 
evaluated at $\boldsymbol{\theta}_0$, 
and $\mathbf\Sigma_\varepsilon(\boldsymbol{\theta}_0)$ is the asymptotic covariance matrix of the empirical gradient
\cite[see, e.g.][]{goldfeld2023statistical,Goldfeld2024}. Hence, $\widehat{\boldsymbol{\theta}}_n^{\mathrm{EOT}}$ satisfies

\begin{equation}\label{eq:EOT_minus_theta_0}
\widehat{\boldsymbol{\theta}}_n^{\mathrm{EOT}} - \boldsymbol{\theta}_0 = O_{\mathbb P}(n^{-1/2}).
\end{equation}

Consequently, our strategy is twofold: we compare $\widehat{\boldsymbol{\theta}}_n^{\mathrm{AW}}$ against $\widehat{\boldsymbol{\theta}}_n^{\mathrm{EOT}}$ at the level of asymptotic rates, and $\widehat{\boldsymbol{\theta}}_n^{\mathrm{AW}}$ against $\widehat{\boldsymbol{\theta}}_n^{\mathrm{NBE}}$ through their discrepancy since the convergence rate of the NBE estimator is currently unknown.

To this end, we introduce the following additional hypotheses:

\begin{enumerate}
\item[$(H4^*)$] (Penalty decay). We replace $(H4)$ with a stronger condition that the penalty sequence $(\lambda_n)_n$ satisfies
$$
\lambda_n \sqrt n \xrightarrow[n\to\infty]{} 0.
$$
This condition ensures that the regularization term in Equation~\eqref{eq:expansion_assumption} is asymptotically negligible compared to the stochastic fluctuations of the empirical Sinkhorn score 
$\nabla S_\varepsilon(\widehat P_n,\widehat P_{m_n,\boldsymbol{\theta}})$, which are of order $n^{-1/2}$ as assumed in the next assumption.
\item[$(H5)$] (Local stochastic expansion of $\nabla S_\varepsilon$).
Motivated by~\eqref{eq:EOT_minus_theta_0}, we assume that there exists a neighborhood 
$\mathcal U$ of $\boldsymbol{\theta}_0$ such that, for any sequence
$(\boldsymbol{\theta}_n)_n$ in $\mathcal U$ satisfying 
$\boldsymbol{\theta}_n \xrightarrow{\mathbb P}\boldsymbol{\theta}_0$,
\begin{equation}\label{eq:expansion_assumption}
\nabla S_\varepsilon(\widehat P_n,\widehat P_{m_n,\boldsymbol{\theta}_n})
=
\nabla S_\varepsilon(\widehat P_n,\widehat P_{m_n,\boldsymbol{\theta}_0})
+
\mathbf V_\varepsilon(\boldsymbol{\theta}_0)(\boldsymbol{\theta}_n-\boldsymbol{\theta}_0)
+
o_{\mathbb P}\bigl(\|\boldsymbol{\theta}_n-\boldsymbol{\theta}_0\|_2 + n^{-1/2}\bigr),
\end{equation}
where $\mathbf V_\varepsilon(\boldsymbol{\theta}_0)$ is a symmetric positive definite matrix.
This assumption should be understood as a local stochastic linearization.
The term $\|\boldsymbol{\theta}_n-\boldsymbol{\theta}_0\|_2$ reflects first-order differentiability with respect to the parameter,
while the $n^{-1/2}$ term accounts for stochastic fluctuations of the empirical measure.
Such fluctuations arise because smooth functionals of empirical measures admit
asymptotically linear expansions and satisfy a central limit theorem,
which implies variations at the $n^{-1/2}$ scale;
see \cite[Chap.~5 and 20]{vandervaart1998}. Assumption $(H5)$ should be read as a model-specific regularity condition
rather than as a consequence of existing results.
Existing limit theorems for the empirical Sinkhorn divergence
\citep{goldfeld2023statistical,Goldfeld2024} describe its behavior when
a target measure is replaced by its empirical counterpart based on a sample
of growing size. They yield a first-order expansion in which a leading
linear term governs the asymptotic distribution and a remainder of smaller
order is negligible at the rate $n^{-1/2}$.
Assumption $(H5)$ is of a different nature: it requires a first-order
expansion in the model parameter $\boldsymbol{\theta}$, along the family
$\{P_{\boldsymbol{\theta}}:\boldsymbol{\theta}\in\Theta\}$, with the simulated
measure $\widehat P_{m_n,\boldsymbol{\theta}}$ replacing $P_{\boldsymbol{\theta}}$.
Establishing such an expansion additionally requires that the model map
$\boldsymbol{\theta} \mapsto P_{\boldsymbol{\theta}}$ be sufficiently smooth
and that the simulation error in $\widehat P_{m_n,\boldsymbol{\theta}}$ be
controlled uniformly in $\boldsymbol{\theta}$.
Existing results make $(H5)$ plausible but do not, to our knowledge, imply it.
Accordingly, Theorem~\ref{thm:asymptotics_AWNBE} should be understood as a
\emph{conditional} asymptotic result: it holds whenever $(H5)$ holds. The
empirical findings of Section~\ref{sec:real_data} do not depend on $(H5)$.
\end{enumerate}
Conditions $(H4^*)$ and  $(H5)$ ensure that the penalty term in the criterion
$$
Q_n(\boldsymbol{\theta})
=
S_\varepsilon(\widehat P_n,\widehat P_{m_n,\boldsymbol{\theta}})
+
\lambda_n \|\boldsymbol{\theta}-\widehat{\boldsymbol{\theta}}_n^{\mathrm{NBE}}\|_2^2
$$
is asymptotically negligible at the $n^{-1/2}$ scale.
Indeed, differentiating the penalty term yields
$$
2\lambda_n(\boldsymbol{\theta}-\widehat{\boldsymbol{\theta}}_n^{\mathrm{NBE}}).
$$
Since $\Theta$ is compact under $(H1)$, both $\boldsymbol{\theta}$ and 
$\widehat{\boldsymbol{\theta}}_n^{\mathrm{NBE}}$ remain bounded, so that 
$\|\boldsymbol{\theta}-\widehat{\boldsymbol{\theta}}_n^{\mathrm{NBE}}\|_2 = O_{\mathbb P}(1)$
and the penalty gradient is of order $O_{\mathbb P}(\lambda_n)$.
Under $(H4^*)$, this term is therefore $o_{\mathbb P}(n^{-1/2})$. The assumptions $(H4^*)$ and $(H5)$ are chosen so that the gradient of $Q_n$ is locally governed by the Sinkhorn term, 
which yields the same limit as for the EOT estimator.\\[1ex]
The following theorem establishes the asymptotic distribution of 
$\widehat{\boldsymbol{\theta}}_n^{\mathrm{AW}}$ and its relationship with the EOT estimator.
Note that $(H4^*)$ replaces $(H4)$ in the assumptions below.
The detailed proof is deferred to Appendix~\ref{app:asymptotics_AWNBE}.
\begin{theorem}\label{thm:asymptotics_AWNBE}
Under $(H1)$--$(H3)$, $(H4^*)$ and $(H5)$, we have
$$
\widehat{\boldsymbol{\theta}}_n^{\mathrm{AW}}
-
\widehat{\boldsymbol{\theta}}_n^{\mathrm{EOT}}
=
o_{\mathbb P}(n^{-1/2}),
$$
and, therefore,
$$
\sqrt n(\widehat{\boldsymbol{\theta}}_n^{\mathrm{AW}}-\boldsymbol{\theta}_0)
\xrightarrow{d}
\mathcal N\!\left(
0,
\mathbf V_\varepsilon(\boldsymbol{\theta}_0)^{-1}
\mathbf\Sigma_\varepsilon(\boldsymbol{\theta}_0)
\mathbf V_\varepsilon(\boldsymbol{\theta}_0)^{-1}
\right), \quad\text{as}\;n\to\infty,
$$
where $\mathbf V_\varepsilon(\boldsymbol{\theta}_0)$ is the Hessian of 
$Q(\boldsymbol{\theta}) = S_\varepsilon(P_{\boldsymbol{\theta}_0}, P_{\boldsymbol{\theta}})$ 
evaluated at $\boldsymbol{\theta}_0$, assumed to be symmetric positive definite,
and $\mathbf\Sigma_\varepsilon(\boldsymbol{\theta}_0)$ is the asymptotic covariance matrix of the empirical gradient as defined in Equation~\eqref{eq:EOT_asymptotics}.
\end{theorem}
Theorem~\ref{thm:asymptotics_AWNBE} proves that the AW--NBE estimator is asymptotically equivalent to the EOT estimator, thereby inheriting the same theoretical guarantees while remaining numerically easier to optimize thanks to the NBE-based regularization. 

\begin{remark}[Conditional nature of Theorem~\ref{thm:asymptotics_AWNBE}]
\label{rmk:H5}
Once $(H5)$ is assumed, the proof of Theorem~\ref{thm:asymptotics_AWNBE}
follows by standard M-estimation arguments \citep{vandervaart1998}. The
novel content of the theorem is therefore not the M-estimation step itself,
but the asymptotic equivalence
$\widehat{\boldsymbol{\theta}}_n^{\mathrm{AW}}
-\widehat{\boldsymbol{\theta}}_n^{\mathrm{EOT}}
=o_{\mathbb P}(n^{-1/2})$,
which shows that the contribution of the NBE-based penalty in
$Q_n$ is asymptotically negligible at the $n^{-1/2}$ scale. Deriving $(H5)$
from explicit smoothness conditions on the parametric family
$\{P_{\boldsymbol{\theta}}:\boldsymbol{\theta}\in\Theta\}$ and on the entropic
regularization parameter $\varepsilon$ remains open, and constitutes the main
theoretical limitation of the present asymptotic analysis. Consequently, the
role of NBE in $\widehat{\boldsymbol{\theta}}_n^{\mathrm{AW}}$ is primarily
finite-sample and computational rather than asymptotic.
\end{remark}
We now compare the AW--NBE estimator with the classical NBE estimator. Since the asymptotic behavior of the NBE estimator has not yet been theoretically characterized, we do not attempt a direct comparison of their asymptotic distributions. Instead, we compare the two estimators through the empirical Sinkhorn discrepancy.
\begin{proposition}
\label{prop:AW_vs_NBE}
Let $(m_n)_{n\ge1}$ be the simulation sample size sequence introduced in Definition~\ref{def:AWNBE_sinkhorn}, and let
$\widehat{\boldsymbol{\theta}}_n^{\mathrm{AW}}$ be as in the same definition. Then, for any positive integer $n$,
$$
S_\varepsilon(\widehat P_n,
\widehat P_{m_n,\widehat{\boldsymbol{\theta}}_n^{\mathrm{AW}}})
\le
S_\varepsilon(\widehat P_n,
\widehat P_{m_n,\widehat{\boldsymbol{\theta}}_n^{\mathrm{NBE}}}).
$$
\end{proposition}
Proposition~\ref{prop:AW_vs_NBE} provides a finite-sample comparison with the NBE and establishes that the AW--NBE estimator always achieves an empirical Sinkhorn discrepancy no larger than that of the NBE estimator; 
see the proof in Appendix~\ref{app:AW_vs_NBE}.
This does not imply a uniform improvement in parameter error, but it shows that the refinement step is effective for the discrepancy of interest. In the empirical results, this refinement appears to have a noticeable impact in finite samples, with a more pronounced effect for discrete distributions than for continuous ones.
\paragraph{An analytical one-dimensional example.}
Before turning to applications, we first consider a simple one-dimensional example where 
the criterion $Q_n(\theta)$ can be computed explicitly.
This example follows the conjugate
model studied in \citet[Section~2.2.3]{sainsbury2024neural}, in particular 
$p=1$ and $d=1$. Let $\{P_\theta:\theta\in\Theta\}$ be a parametric
family of probability distributions on $\mathbb{R}$.
In this example, we take
\[
P_\theta=\mathrm{Unif}(0,\theta), \qquad \Theta\subset\mathbb{R}_+ ,
\]
and a sample
$
X_1,\dots,X_n \mid \theta \ \overset{\text{i.i.d.}}{\sim}\ P_\theta 
$
with prior distribution on $\theta$, 
\[
\theta \sim \mathrm{Pareto}(\alpha,\beta),
\qquad
\alpha>0,\ \beta>0,
\]
where the Pareto distribution with shape parameter $\alpha$ and scale parameter $\beta$ 
is defined by its density
\[
\pi(\theta)
=
\alpha \beta^\alpha \theta^{-(\alpha+1)} \mathds{1}_{(\theta \ge \beta)}.
\]
Under absolute loss $L(\theta,a)=|\theta-a|$, 
the Bayes estimator is the posterior median and admits the closed--form
expression \cite[see, e.g.][Eq.~(8)]{sainsbury2024neural}
\begin{equation}\label{eq:theta_Bayes_example}
\widehat\theta_n^{\mathrm{Bayes}}(X)
=
2^{1/(\alpha+n)}
\max\bigl(X_1,\dots,X_n,\beta\bigr).
\end{equation}

An NBE is an estimator of the form $\widehat{\theta}_n^{\mathrm{NBE}}=\psi(X)$, where 
$\psi$ is obtained by minimizing a Monte Carlo approximation of
the Bayes risk \citep{sainsbury2024neural}. In this example,
$\widehat{\theta}_n^{\mathrm{NBE}}$ targets the Bayes decision rule, i.e.
\begin{equation*}
\widehat{\theta}_n^{\mathrm{NBE}} \approx \widehat{\theta}_n^{\mathrm{Bayes}}
\text{ defined in Equation~\eqref{eq:theta_Bayes_example}}.
\end{equation*}

To make the AW--NBE refinement mechanism explicit, we consider a population
analogue of the criterion introduced in Definition~\ref{def:AWNBE_sinkhorn},
in which the empirical Sinkhorn discrepancy is replaced by its population counterpart
and the NBE by its target. This oracle construction does not define a feasible estimator,
since it depends on the unknown true parameter $\theta_0$, but it illustrates the 
shrinkage mechanism underlying Theorems~\ref{thm:consistency_aw_nbe} and~\ref{thm:asymptotics_AWNBE}.

Fix $\lambda>0$ and define
\[
\bar\theta_{\mathrm{AW}}
\in
\arg\min_{\theta\in\Theta}
\left\{
S_\varepsilon(P_{\theta_0},P_\theta)
+
\lambda\bigl(\theta-\widehat{\theta}_n^{\mathrm{Bayes}}\bigr)^2
\right\}.
\]
For the quadratic cost $c(x,y)=(x-y)^2$, the Sinkhorn divergence $S_\varepsilon(P_{\theta_0},P_\theta)$
converges to $W_2^2(P_{\theta_0},P_\theta)$ as $\varepsilon\downarrow 0$ 
\cite{Feydy2019}. For the chosen model $P_\theta=\mathrm{Unif}(0,\theta)$, 
the quantile function is $F_\theta^{-1}(u)=\theta u$ for any $u\in[0,1]$.
Hence the squared $2$-Wasserstein distance between $P_{\theta_0}$ and $P_\theta$ is
\[
W_2^2(P_{\theta_0},P_\theta)
=
\int_0^1
\bigl(F_{\theta_0}^{-1}(u)-F_\theta^{-1}(u)\bigr)^2\,du
=
\int_0^1 (\theta_0 u-\theta u)^2\,du
=
\frac{(\theta-\theta_0)^2}{3}.
\]
Replacing $S_\varepsilon$ by this limit yields the oracle objective
\[
J(\theta)
=
\frac{(\theta-\theta_0)^2}{3}
+
\lambda\bigl(\theta-\widehat{\theta}_n^{\mathrm{Bayes}}\bigr)^2.
\]
Since $J$ is strictly convex, its minimizer is
\begin{equation*}
\bar\theta_{\mathrm{AW}}
=
\frac{\theta_0/3+\lambda\,\widehat{\theta}_n^{\mathrm{Bayes}}}
{\lambda+1/3},
\end{equation*}
which can be rewritten as a weighted average of $\theta_0$ and 
$\widehat{\theta}_n^{\mathrm{Bayes}}$:
\begin{equation*}
\bar\theta_{\mathrm{AW}}
=
\frac{1}{1+3\lambda}\,\theta_0
+
\left(1-\frac{1}{1+3\lambda}\right)\widehat{\theta}_n^{\mathrm{Bayes}}.
\end{equation*}
Substituting the closed-form expression~\eqref{eq:theta_Bayes_example} gives
\begin{equation*}
    \bar\theta_{\mathrm{AW}}
=
\frac{1}{1+3\lambda}\,\theta_0
+
\left(1-\frac{1}{1+3\lambda}\right)2^{1/(\alpha+n)}
\max(X_1,\dots,X_n,\beta).
\end{equation*}
This provides a finite-sample illustration of the refinement mechanism in 
Theorems~\ref{thm:consistency_aw_nbe} and~\ref{thm:asymptotics_AWNBE}. The transport term shrinks the estimate toward $\theta_0$, while $\lambda$ controls the relative weight of the NBE initialization. The feasible AW--NBE estimator $\widehat{\theta}_n^{\mathrm{AW}}$ replaces $S_\varepsilon(P_{\theta_0}, P_\theta)$ with the empirical Sinkhorn discrepancy 
$S_\varepsilon(\widehat{P}_n, \widehat{P}_{m_n,\theta})$ and $\widehat{\theta}_n^{\mathrm{Bayes}}$ with $\widehat{\theta}_n^{\mathrm{NBE}}$, thereby recovering the criterion of Definition~\ref{def:AWNBE_sinkhorn}.\\[1ex]

We now describe the metrics used to assess goodness-of-fit, which will be applied in Section~\ref{sec:real_data}.

\subsection{Goodness-of-fit assessment}
\label{sec:gof}
To assess the goodness-of-fit of the fitted models, we consider the null hypothesis $H_0$ that the true data-generating distribution belongs to a chosen parametric family 
$$
\{P_{\boldsymbol{\theta}} : \boldsymbol{\theta} \in \Theta\}.
$$
We test this hypothesis using the Sinkhorn divergence introduced in Equation~\eqref{eq:Sinkhorn}, calibrated via a parametric bootstrap procedure \cite[see e.g.,][]{EfronTibshirani1994,vandervaart1998}.

Let $\widehat P_n$ and $\widehat P_{m_n,\widehat{\boldsymbol{\theta}}_n}$ 
denote the empirical measures defined in 
Equation~\eqref{eq:empirical_measures}, where $(m_n)_{n\geq 1}$ 
is a sequence of positive integers with $m_n \to \infty$ as 
$n \to \infty$. These are based, respectively, on the observed sample $\boldsymbol{X}_{n,d}$ and on a sample of size $m_n$ simulated from the fitted model $P_{\widehat{\boldsymbol{\theta}}_n}$. We define the observed discrepancy as
$$
D_n(\widehat{\boldsymbol{\theta}}_n)
=
S_\varepsilon\!\left(
\widehat P_n,
\widehat P_{m_n,\widehat{\boldsymbol{\theta}}_n}
\right),
$$
where smaller values indicate a better agreement between the data and the fitted model.

The goal is to assess whether the observed discrepancy $D_n(\widehat{\boldsymbol{\theta}}_n)$ is larger than what would be expected due to finite-sample variability. In our setting, the chosen parametric family lacks a closed-form likelihood, and the finite-sample distribution of the Sinkhorn discrepancy is analytically intractable while depending on the unknown true parameter. We therefore approximate this distribution using a parametric bootstrap. In this approach, we simulate replicated datasets under the fitted model $P_{\widehat{\boldsymbol{\theta}}_n}$, which serves as a plug-in estimate of the true distribution. 

The procedure is carried out as follows: For $b=1,\dots,B$:
\begin{enumerate}[label=(\roman*)]
    \item Generate a parametric bootstrap sample 
    $\boldsymbol{X}_1^{(b)},\dots,\boldsymbol{X}_n^{(b)} \sim P_{\widehat{\boldsymbol{\theta}}_n}$, 
    and denote by $\widehat P_n^{(b)}$ its empirical distribution.
    \item Estimate the model parameters on this bootstrap sample 
    to obtain $\widehat{\boldsymbol{\theta}}_n^{(b)}$.
    \item Simulate a new sample of size $m_n$ from the re-fitted model 
    $P_{\widehat{\boldsymbol{\theta}}_n^{(b)}}$, with empirical measure 
    $\widehat P_{m_n,\widehat{\boldsymbol{\theta}}_n^{(b)}}$.
    \item Compute the parametric bootstrap discrepancy:
    $$
    D_b = S_\varepsilon\!\left(
    \widehat P_n^{(b)},
    \widehat P_{m_n,\widehat{\boldsymbol{\theta}}_n^{(b)}}
    \right).
    $$
\end{enumerate}

Finally, the bootstrap $p$-value is calculated as (see \citet{EfronTibshirani1994}):
$$
p\text{-value} = \frac{1}{B+1} \left( 1 + \sum_{b=1}^B \mathds{1}_{\left\{ D_b \ge D_n(\widehat{\boldsymbol{\theta}}_n) \right\}} \right);
$$
it evaluates whether the observed discrepancy is unusually large compared to the discrepancies expected under the fitted model. Large $p$-values indicate that the model provides an adequate fit, whereas small $p$-values suggest a lack of fit. A formal validity analysis of this parametric bootstrap calibration for the Sinkhorn discrepancy under simulation-based estimation is beyond the scope of the present paper and is left for future work. The procedure should therefore be interpreted as a practical diagnostic and model-adequacy tool rather than as a hypothesis test with established asymptotic guarantees in this setting.
\\[1ex]
Beyond parameter estimation, EOT provides graphical diagnostic tools for model validation. 
Following \citet{singha2024}, we use EOT maps and potentials to visually assess model adequacy.

Let $\mu$ denote the uniform distribution on the unit ball $B^d\subset\mathbb{R}^d$, and let $\boldsymbol{U}_1,\dots,\boldsymbol{U}_n
\overset{\mathrm{iid}}{\sim}\mu$ be auxiliary reference points, independent of the observed and simulated samples.

For $\displaystyle \widehat{P}\in \{\widehat{P}_n,\widehat{P}_{m_n,\widehat{\boldsymbol{\theta}}_n^{\mathrm{Method}}}\}$,
the EOT map, more precisely the barycentric projection of the entropic coupling,
$T^\varepsilon_{\widehat{P}}:B^d\to\mathbb{R}^d$ and the source-side EOT potential $\phi^\varepsilon_{\widehat{P}}:B^d\to\mathbb{R}$
are defined through the entropy-regularized optimal transport problem from $\mu$ to $\widehat{P}$ under the quadratic cost
$\displaystyle c(\boldsymbol{u},\boldsymbol{x}) =\frac12\|\boldsymbol{u}-\boldsymbol{x}\|^2$, 
that is,
\[ 
T^\varepsilon_{\widehat{P}}(\boldsymbol{u}) =
\mathbb{E}_{\pi^\star} [\boldsymbol{X}\mid\boldsymbol{U}=\boldsymbol{u}],
\qquad \text{where}\; (\boldsymbol{U},\boldsymbol{X})
\sim \pi^\star_{\mu,\widehat{P}},
\]
where $\pi^\star_{\mu,\widehat{P}}$ denotes the optimal entropic coupling
between $\mu$ and $\widehat{P}$.
Under the quadratic cost, the barycentric projection admits the representation
\[
T^\varepsilon_{\widehat{P}}(\boldsymbol{u}) = \boldsymbol{u} - \nabla\phi^\varepsilon_{\widehat{P}}(\boldsymbol{u}),
\qquad \boldsymbol{u}\in B^d,
\]
see \citet[Section~3]{pooladian2021}.
The two transports compared in the diagnostics share the same source measure:
\[
\mu\longrightarrow\widehat{P}_n, \qquad \mu\longrightarrow \widehat{P}_{m_n,\widehat{\boldsymbol{\theta}}_n^{\mathrm{Method}}},
\]
which makes it possible to compare the resulting transports at common reference locations $\boldsymbol{U}_j$.

The EOT Q--Q plot in coordinate $i$ is
\[
\Delta_i^\varepsilon = \Bigl\{ \bigl(\langle T^\varepsilon_{\widehat{P}_n}(\boldsymbol{U}_j), \boldsymbol{e}_i\rangle,\,\langle T^\varepsilon_{
\widehat{P}_{m_n,\widehat{\boldsymbol{\theta}}_n^{\mathrm{Method}}}}(\boldsymbol{U}_j), \boldsymbol{e}_i \rangle \bigr): j=1,\dots,n \Bigr\},
\]
where $\boldsymbol{e}_i$ is the $i$-th canonical basis vector of $\mathbb{R}^d$.
If the fitted model adequately reproduces the data distribution, the points in
$\Delta_i^\varepsilon$ should concentrate around the diagonal.
The EOT potential plot is
\[
\Gamma^\varepsilon = \Bigl\{ \bigl( \phi^\varepsilon_{\widehat{P}_n}(\boldsymbol{U}_j),\, \phi^\varepsilon_{ \widehat{P}_{m_n,\widehat{\boldsymbol{\theta}}_n^{\mathrm{Method}}}}(\boldsymbol{U}_j)\bigr): j=1,\dots,n\Bigr\}.
\]
Concentration around the diagonal indicates similar transport structures for the observed and fitted distributions.
A practical advantage of the potential plot is that it provides a single diagnostic representation regardless of the dimension $d$.

To statistically validate these visual assessments, \citet{singha2024} introduced the discrepancy statistics
\begin{equation}\label{eq:eot_pvalue}
E_n^{\varepsilon} = n\int_{B^d}\bigl\|T^\varepsilon_{\widehat{P}_n}(\boldsymbol{u})-T^\varepsilon_{\widehat{P}_{m_n,\widehat{\boldsymbol{\theta}}_n^{\mathrm{Method}}}}(\boldsymbol{u})\bigr\|^2\,d\mu(\boldsymbol{u}),
\quad F_n^{\varepsilon}=n\int_{B^d}\bigl|\phi^\varepsilon_{\widehat{P}_n}(\boldsymbol{u})-\phi^\varepsilon_{\widehat{P}_{m_n,\widehat{\boldsymbol{\theta}}_n^{\mathrm{Method}}}}(\boldsymbol{u})\bigr|^2\,d\mu(\boldsymbol{u}),
\end{equation}
and developed formal goodness-of-fit tests based on these quantities
(see Theorem~4.2 therein), relying on functional central limit theorems for empirical EOT maps and potentials established by \citet{Goldfeld2024}.
\paragraph{Adaptation to the discrete case for dry spells data.}
Since the dry spell exceedances take integer values, many observations are identical. Plotting all $n$ data points individually would lead to overlapping points, hiding the true concentration of the data. 

To avoid this and properly reflect the discrete structure in the EOT diagnostics, we represent the empirical measure by its distinct support points $\boldsymbol{z}_1,\dots,\boldsymbol{z}_m$ together with their empirical frequencies $p_1,\dots,p_m$, so that
\[
\widehat{P}_n = \sum_{k=1}^m p_k\,\delta_{\boldsymbol{z}_k}.
\]
In the EOT Q--Q and potential plots, each distinct value is plotted once, with a marker size proportional to its empirical frequency $p_k$. This weighted representation avoids overplotting and makes the discrete probability mass visually explicit, while remaining consistent with the general EOT framework of \citet{singha2024}.

Regarding the entropic regularization parameter $\varepsilon$, it is well known that it dictates the trade-off between approximating the unregularized Wasserstein distance (small $\varepsilon$) and ensuring numerically stable, smooth transport maps (large $\varepsilon$) \citep{cuturi2013sinkhorn,Feydy2019}. In our setting, we found the diagnostics to be qualitatively stable across a wide range of regularization levels. Consequently, we fix $\varepsilon = 10^{-2}$ in all figures, as it provides an effective compromise between computational stability and visual interpretability.

\section{Application to financial and environmental time series}
\label{sec:real_data}

In this section, we first revisit the financial example of \citet{Kiriliouk2019}, which analyzes weekly negative log-returns of four major U.K. banks over 2007–2016, keeping their thresholding and generator $\boldsymbol{T}$ (Gumbel–$\boldsymbol{T}$ MGPD model): we use this threshold model as a benchmark to compare censored likelihood estimation with likelihood-free estimators based on neural Bayes inference and Wasserstein refinements. The financial dataset provides a controlled benchmark in which model specification and threshold choices are kept from the previous study, allowing us to isolate the impact of the inference methodology.
Second, we turn to discrete environmental data and apply the MDGPD framework to dry spell lengths from Swiss precipitation records. 
Here, parameters are estimated either by AW--NBE or via NBE trained on simulated MDGPD samples. 
Across both continuous and discrete settings, the goal is not to re-select models (i.e. to propose a new MGPD model) but to show, on real data, how OT-based and likelihood-free approaches perform relative to the classical censored likelihood, using the same thresholds and generators as in the original studies.

\subsection{Banks log-returns}

The empirical illustration in \cite{Kiriliouk2019} considers weekly negative returns of four major U.K. banks, denoted $(X_{t}^{(1)},X_{t}^{(2)},X_{t}^{(3)},X_{t}^{(4)})$ for HSBC, Lloyds, RBS, and Barclays respectively, over the period from 2007/10/29 to 2016/10/17 ($n=470$ observations). Their focus is on modeling joint tail events,
\[
\bigcup_{i=1}^{4}\,\bigl\{X_{t}^{(i)} > u_{i}\bigr\},
\]
where $u_{i}$ denotes a high marginal threshold, chosen at the $83^{\text{rd}}$ percentile.

After marginal standardization to exponential margins via the probability integral transform, 
excesses are extracted and rescaled to obtain observations in the canonical domain
\[
\{ z \in \mathbb{R}_{+}^{4} : \max_{1 \le j \le 4} z_{j} > 1 \}.
\]
Multivariate extreme value theory implies that the MGPD arises as an asymptotic model for exceedances in the joint tail region; see \cite{rootzen_segers_wadsworth} (or, e.g., the survey by Naveau and Segers in \cite[Chap.~7]{HandbookExtremes2026}). The density of the MGPD depends on the distribution of a generator vector $\boldsymbol{T}$. 
We recall that a standard MGPD vector $\mathbf{Z}$ can be represented as $\mathbf{Z} = E\,\mathbf{1} + \mathbf{S}$, where $E$ is a unit exponential random variable independent of $\mathbf{S}$, with $\max \mathbf{S} = 0$; see Appendix~\ref{sec:mgpds_def} for further details.
The non-positive vector $\boldsymbol{T}=\boldsymbol{S}-\max \boldsymbol{S}$ is called a generator.
In the standardized case $(\sigma=1,\gamma=0)$,
if $\boldsymbol{T}$ has density $f_{\boldsymbol{T}}$, the MGPD density can be written as
\[
\mathds{1}_{\{\max(x)>0\}} e^{\max(x)} \int_0^\infty f_{\boldsymbol{T}}(x+\log t)\,t^{-1}dt ,
\]
see Equation~(3.3) in \cite{Kiriliouk2019}. 
Different choices of the generator distribution $\boldsymbol{T}$ result in distinct MGPD families. Examples considered in \cite{Kiriliouk2019} include constructions based on Gumbel, Gaussian, and reverse exponential distributions.
Parameters were estimated by maximizing a censored likelihood (henceforth referred to as MLE), with model selection performed using the Akaike Information Criterion (AIC) and likelihood ratio tests. Ultimately, the Gumbel-$T$ family was identified as providing the best fit for the bank data.

\begin{figure}[H]
\centering
\includegraphics[width=0.65\linewidth]{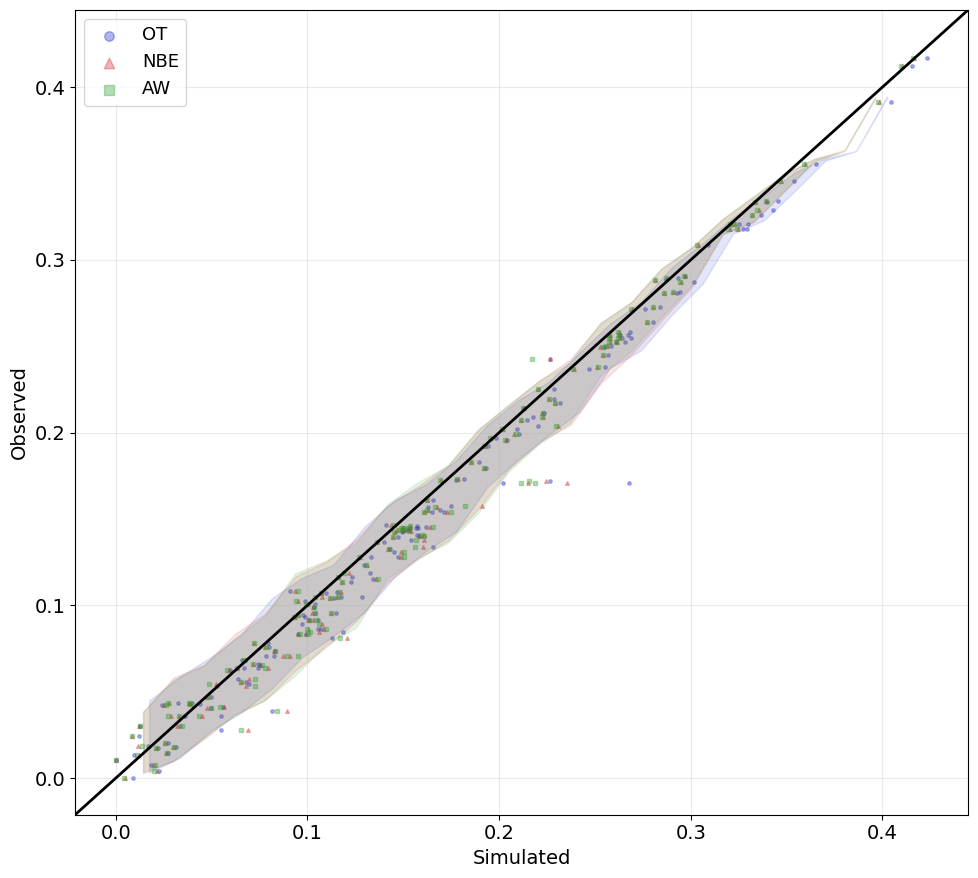}
\caption{\small \sf EOT potential plots for the bank log-return exceedances dataset under the MGPD model.
The three estimation procedures are displayed: OT (grey circles), NBE (red triangles), and AW--NBE (green squares).}
\label{fig:PotentialBanksTransportComparison}
\end{figure}

\begin{figure}[H]
\centering
\includegraphics[width=0.9\linewidth]{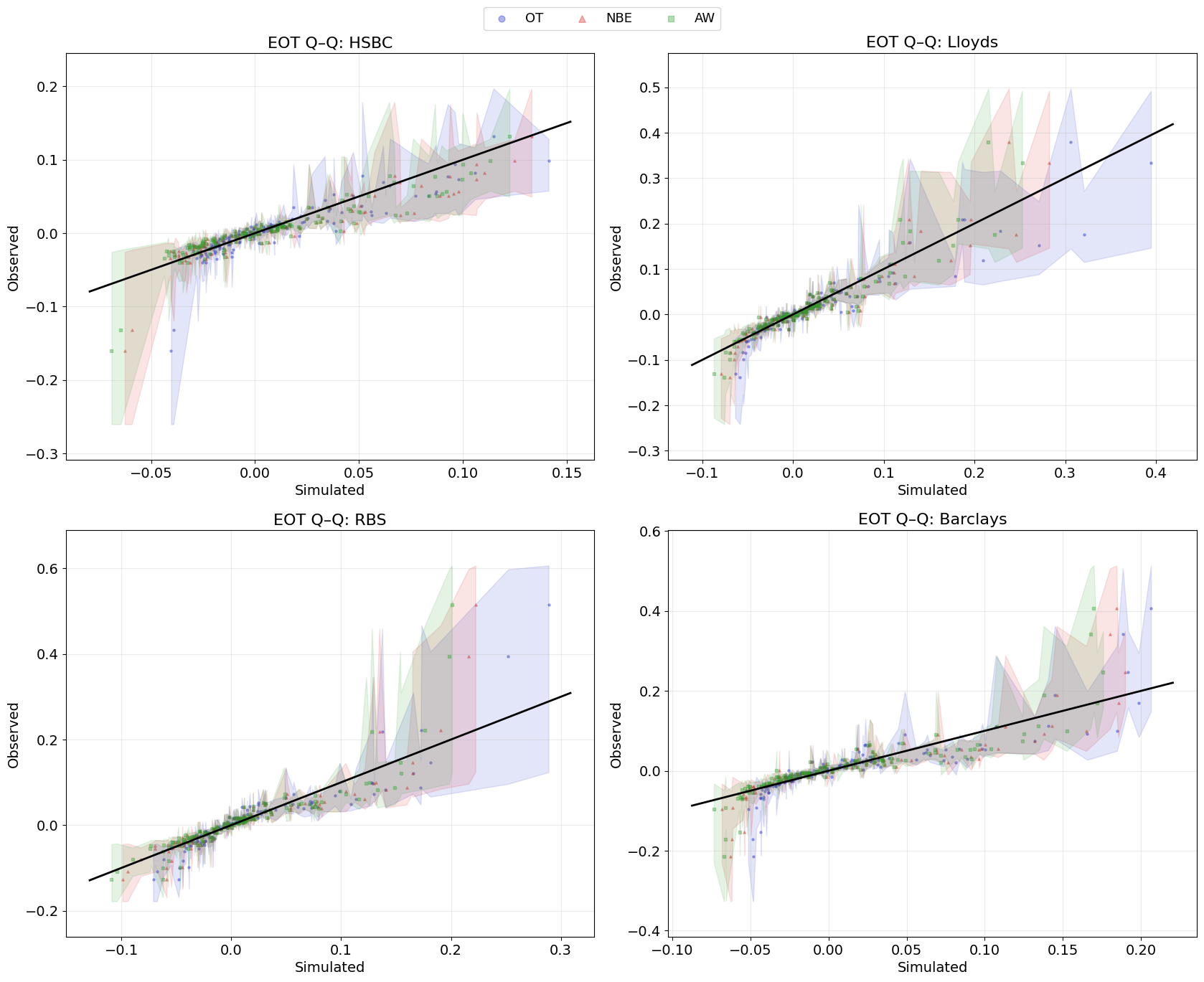}
\caption{\small \sf
Quadrivariate EOT Q--Q plots for the bank log-return exceedances dataset under the MGPD model.
The plots compare the OT in grey circles, the NBE in red triangles, and the AW--NBE in green squares.
Each panel corresponds to the projection onto the $e_i$ coordinate, $i=1,\ldots,4$, where $e_1$, $e_2$, $e_3$, and $e_4$ correspond respectively to HSBC, Lloyds, RBS, and Barclays.
}
\label{fig:QQplotBanksTransportComparison}
\end{figure}

In what follows, we do not revisit their model selection. 
We retain their generator, namely the Gumbel--$\boldsymbol{T}$ MGPD, together with the same threshold choice ($83^{\text{rd}}$ percentile) and censoring framework. Our objective is purely comparative: we contrast estimation \emph{criteria} while keeping the model parametric specification fixed. All conclusions are therefore conditional on the Gumbel--$\boldsymbol{T}$ MGPD model class.
\\[2ex] 
Applying the goodness-of-fit scheme of Section~\ref{sec:gof} yields Table~\ref{tab:gof_banks}, which reports $D_n(\widehat{\boldsymbol{\theta}}_n)$ and the corresponding bootstrap $p$-values for four different estimation methods.
\begin{table}[H]
\centering
\caption{\small \sf Goodness-of-fit discrepancies and bootstrap $p$-values for the banking 
loss dataset. Smaller values of $D_n(\widehat{\boldsymbol{\theta}}_n)$ indicate 
better agreement between the observed and simulated joint exceedances. For each method, 
$D_n(\widehat{\boldsymbol{\theta}}_n) = S_\varepsilon(\widehat{P}_n, 
\widehat{P}_{m_n, \widehat{\boldsymbol{\theta}}_n})$, where 
$\widehat{\boldsymbol{\theta}}_n \in \{\widehat{\boldsymbol{\theta}}_n^{\mathrm{AW}}, 
\widehat{\boldsymbol{\theta}}_n^{\mathrm{NBE}}, 
\widehat{\boldsymbol{\theta}}_n^{\mathrm{EOT}},
\widehat{\boldsymbol{\theta}}_n^{\mathrm{MLE}}\}$. The quantity $|\Delta|$ denotes the percentage reduction in discrepancy relative to MLE.}
    \begin{tabular}{lccc}
    \hline
    Method & $D_n(\widehat{\boldsymbol{\theta}}_n)$ & $|\Delta|$ & Bootstrap $p$-value \\
    \hline
    AW--NBE & $0.0189$ & $36.8\%$ & $0.20$\\
    NBE     & $0.0209$ & $30.1\%$ & $0.11$ \\
    OT      & $0.0246$ & $17.7\%$ & $0.17$ \\
    MLE     & $0.0299$ & $-$ & $0.21$      \\
    \hline
    \end{tabular}
\label{tab:gof_banks}
\end{table}
\paragraph{Caveat on evaluation.}
Since the goodness-of-fit discrepancy $D_n$ is itself based on the Sinkhorn
divergence, estimators that explicitly minimize Sinkhorn-type criteria
(OT and AW--NBE) are favored by construction over those that do not
(MLE, NBE). The comparison nevertheless remains informative because all
estimators are evaluated on the same fitted model class with identical
thresholds, simulation size, and entropic regularization parameter; we
therefore interpret it as a comparison under a common OT-based diagnostic
rather than as evidence of universal superiority.\\[1ex]
The AW--NBE estimator achieves the smallest discrepancy ($0.0189$), followed by NBE ($0.0209$), OT ($0.0246$), and MLE ($0.0299$). Although these results indicate a moderate global improvement, it is important to note that the Wasserstein-based metric $D_n$ aggregates agreement across the entire support. Consequently, refinements achieved in the tail regions may be partially smoothed out in this global average. Nevertheless, all bootstrap $p$-values are above standard significance levels, indicating that none of the fitted models is rejected by the goodness-of-fit test. Among the four estimators considered here, AW--NBE achieved the smallest discrepancy, followed by NBE, OT, and MLE.\\[1ex]
In Figure~\ref{fig:PotentialBanksTransportComparison}, we present the EOT potential plots. While NBE and OT display broadly similar patterns, with only moderate differences, the AW--NBE estimator exhibits the strongest concentration along the diagonal. This suggests that the Wasserstein refinement improves the overall agreement between the empirical and simulated distributions, further supporting the results in Table~\ref{tab:gof_banks}.
In Figure~\ref{fig:QQplotBanksTransportComparison}, we compare the EOT Q--Q plots for the OT, NBE, and AW--NBE estimators. 
All methods effectively capture the global dependence structure, as shown by the overall diagonal alignment, but systematic differences emerge in the upper tail. The OT and NBE estimators exhibit comparable behavior, with moderate deviations across margins. In contrast, the AW--NBE estimator shows a consistently tighter concentration around the diagonal. This improvement is particularly visible in the extreme quantiles, where deviations are noticeably smaller than for both OT and NBE. Visually, the dispersion around the diagonal appears reduced in the upper quantiles, although the bootstrap confidence bands overlap and no formal tail-specific test is performed.
Overall, among the three likelihood-free estimators considered here, AW--NBE achieved the smallest discrepancy, followed by NBE and OT.

\begin{figure}[H]
\centering
\includegraphics[width=0.52\linewidth]{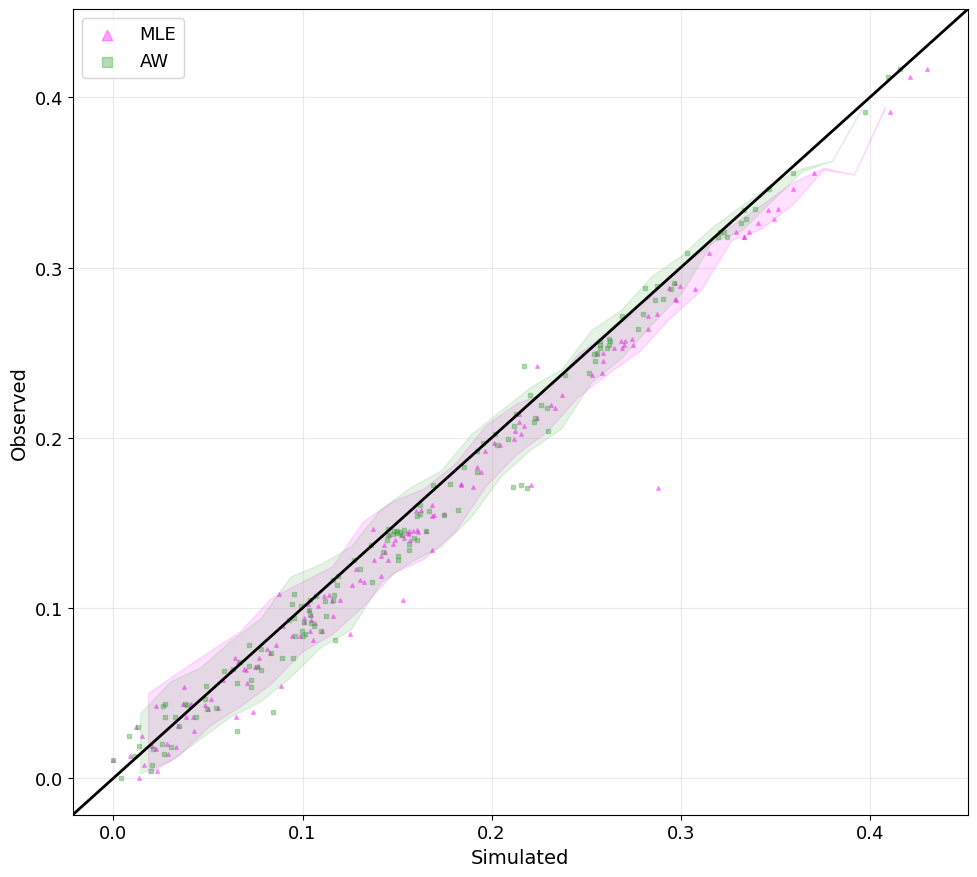}
\caption{\small \sf EOT potential plots for the bank log-return exceedances dataset under the MGPD model.
MLE (blue triangles) and AW--NBE (green squares).}
\label{fig:PotentialBanksTransportComparison_MLE_AW}
\end{figure}
\vspace{-2ex}

\begin{figure}[H]
\centering
\includegraphics[width=0.99\linewidth]{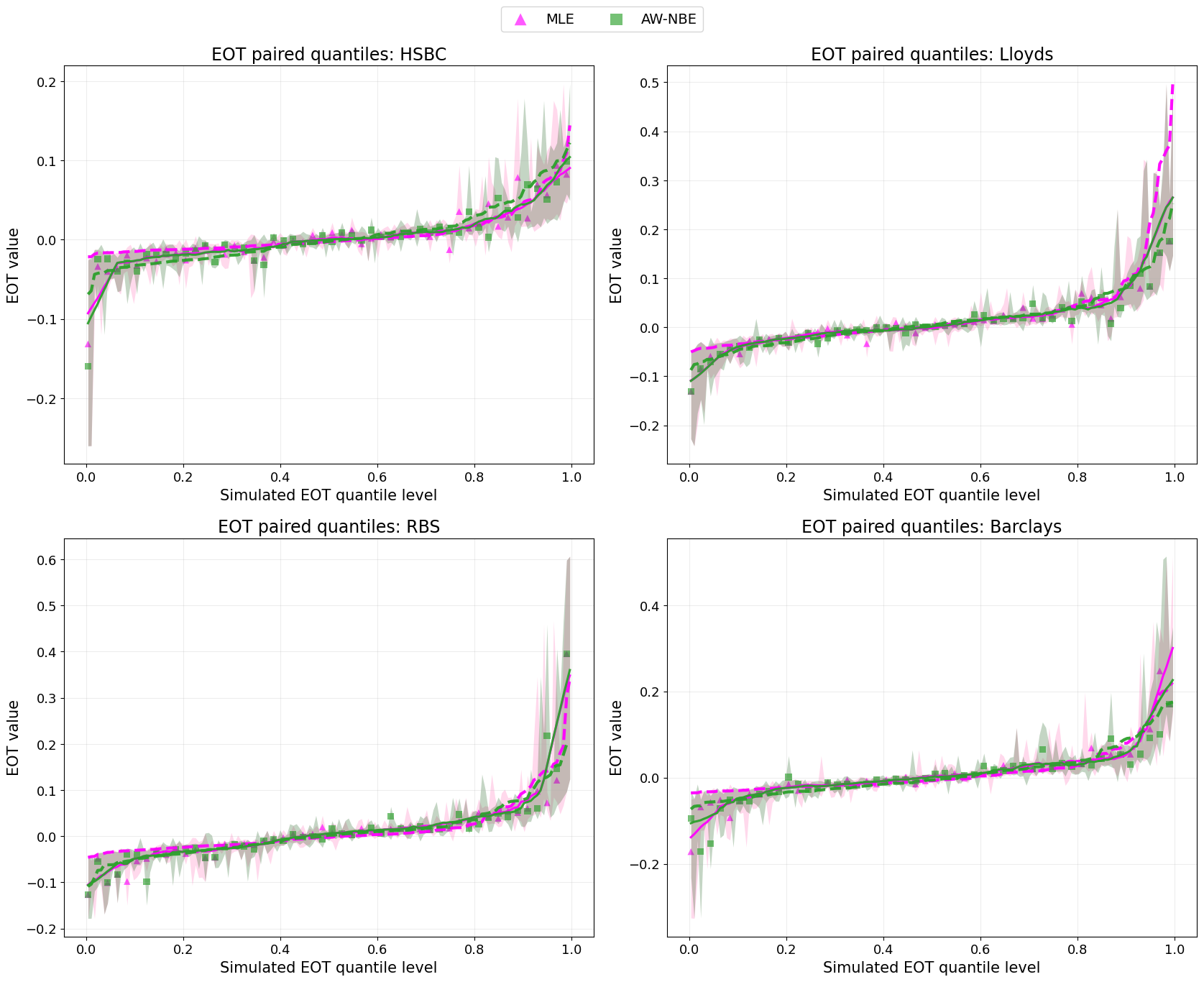}
\caption{\small \sf EOT paired quantile–rank plots for the bank log-return exceedances dataset under the MGPD model. The solid curve shows the observed EOT map, the dashed curve the simulated model EOT map, and shaded bands denote pointwise 95\% bootstrap confidence intervals (pink for MLE, dark green for AW--NBE).}
\label{fig:EOTRANKSMLEAWNBE}
\end{figure}

After analyzing the likelihood-free estimators, we compare them to the censored maximum likelihood estimator (hereafter MLE), introduced in \cite{Kiriliouk2019}, which serves as a parametric benchmark for MGPD models.
For clarity, we focus on a visual comparison between the MLE and AW--NBE estimators with potential (Figure~\ref{fig:PotentialBanksTransportComparison_MLE_AW}) and the EOT paired quantile rank plots (Figure~\ref{fig:EOTRANKSMLEAWNBE}). The AW--NBE serves as a representative likelihood-free method, as the different likelihood-free estimators exhibit broadly similar behavior in the graphical diagnostics.

Figure~\ref{fig:PotentialBanksTransportComparison_MLE_AW} displays the EOT potential plots for the banks dataset under the MLE and AW--NBE estimations. Overall, both methods capture the main dependence structure of the joint exceedances reasonably well, as most points remain concentrated around the identity line. 
However, the AW--NBE estimator exhibits a tighter alignment with the diagonal over a broad range of quantile levels, particularly in the intermediate and upper regions of the distribution. 
The confidence bands further highlight these differences. 
The AW--NBE bands remain globally closer to the diagonal and appear slightly more stable across the support, whereas the MLE fit shows larger deviations in several regions, especially for moderate-to-large simulated potential values. 
These visual diagnostics seem consistent with the discrepancy values reported in Table~\ref{tab:gof_banks}, where AW--NBE achieves the smallest goodness-of-fit discrepancy among all competing estimators.
In Figure~\ref{fig:EOTRANKSMLEAWNBE}, the $x$-axis represents the EOT quantile level, obtained for each estimator by computing EOT-transported values and ordering them component-wise. Although the observed EOT-transported values are identical across estimators, they are displayed according to the empirical rankings induced by the corresponding fitted model. A well-fitted model is characterized by close agreement between the observed (solid) and simulated (dashed) curves, with the observed curve remaining within the bootstrap confidence bands.  

Figure~\ref{fig:EOTRANKSMLEAWNBE} shows that MLE and AW--NBE produce similar results over the central quantile range for all four banks. In the upper tail, MLE tends to produce higher extrapolated quantiles than AW--NBE, most visibly for Lloyds, where the dashed MLE curve clearly exceeds the AW--NBE curve and departs from the observed solid curve. For RBS and HSBC, both methods remain broadly consistent, though MLE shows slightly more variability in the tails. Barclays is the most stable case, with both estimators remaining close throughout.\\[1ex]

While these visual conclusions could be formally validated using the statistical tests defined in \eqref{eq:eot_pvalue}, the bootstrap $p$-values reported in Table~\ref{tab:gof_banks} already provide sufficient evidence for this assessment. This is consistent with the use of likelihood-free approaches in settings where the likelihood may be less accurate due to model or data constraints.

\subsection{Daily dry spells records in Switzerland}

Here, we consider the dry spell dataset constructed from daily precipitation measurements collected by MeteoSwiss \cite[see, e.g.][] {aka2025multivariatediscretegeneralizedpareto} and located at  the two nearby  stations called Interlaken and Lauterbrunnen.  
A dry spell is defined as a sequence of consecutive days with little or no precipitation, and the data consist of the lengths of these runs over several decades of observations in Switzerland (typically 1930–2014). Extreme events are defined as dry spells whose duration exceeds the local 99\textsuperscript{th} percentile at either of the two sites. The use of a high percentile threshold is
consistent with the literature on these hydrological risks \cite[see, e.g.][]{DOMMANGE2026}. As illustrated in \cite{aka2025multivariatediscretegeneralizedpareto}, these lengths are discrete, heavy-tailed, and may occur jointly across nearby stations.
This discrete and dependent structure makes the dataset a natural case study for evaluating the robustness of our inference procedures. 
In the present application, the data consist of time series of dry spell lengths and therefore exhibit temporal dependence. To account for this feature, we use as MDGPD generator a bivariate variable-length Markov process \cite[see, e.g.][]{Rissanen1983,Buhlmann1999,Galves2008}, which provides a way to capture time dependence. Therefore, we use as generator a bivariate variable-length Markov chain (VLMC). 
We briefly recall its definition.
Let $\{X_t\}_{t\in\mathbb{Z}}$ be a stochastic process taking values in a finite or 
countable alphabet $\mathcal{A}$. A VLMC, introduced by 
\citet{Rissanen1983}, is a process for which the conditional distribution of $X_t$ 
given the past depends on a context whose length may vary with the observed history.

More precisely, there exists a set of contexts $\mathcal{C}$ such that
\[
\mathbb{P}(X_t = x \mid X_{t-1},X_{t-2},\dots)
=
p(x \mid c_t),
\]
where $c_t$ denotes the relevant suffix of the past belonging to $\mathcal{C}$.

In the present application, $X_t$ represents the vector of dry spell lengths observed 
at two nearby stations (e.g., Interlaken and Lauterbrunnen). This formulation allows the dependence on the past to adapt to persistence patterns of different durations, which is particularly suitable for modeling dry spell dynamics. The estimator is trained on large sets of data simulated from the MDGPD model, using a DeepSets architecture to guarantee permutation invariance across replicated samples. 
Once trained, the network quickly produces parameter estimates $\boldsymbol{\theta}$ for new datasets, with uncertainty quantified via bootstrap. 
\begin{figure}[H]
\centering
\includegraphics[width=0.7\linewidth]{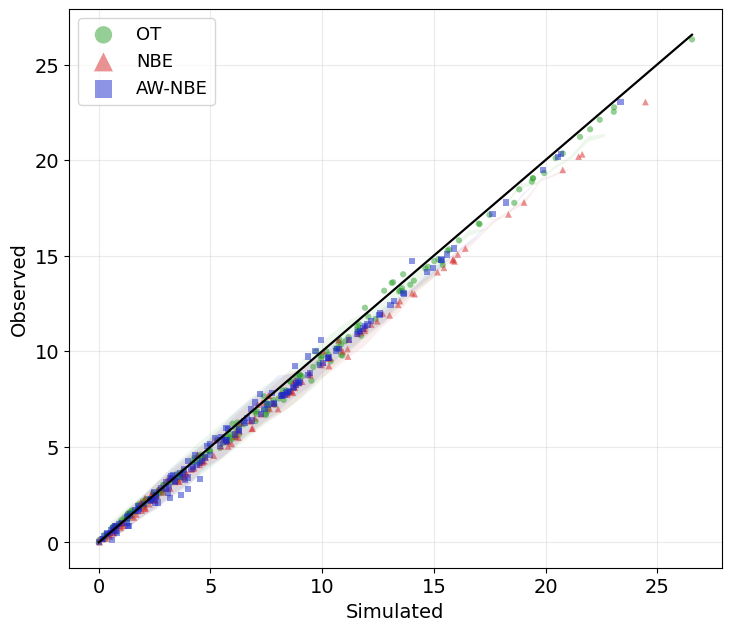}
\caption{\small\sf
EOT potential plots for the dry spell exceedances dataset under the MDGPD model.
The three fitted procedures are displayed OT in green circles, NBE in red triangles and AW--NBE in blue squares.
}
\label{fig:PotentialDrySpellsComparison}
\end{figure}
\begin{figure}[H]
\centering
\hspace{-5ex}
\includegraphics[width=1.03\linewidth]{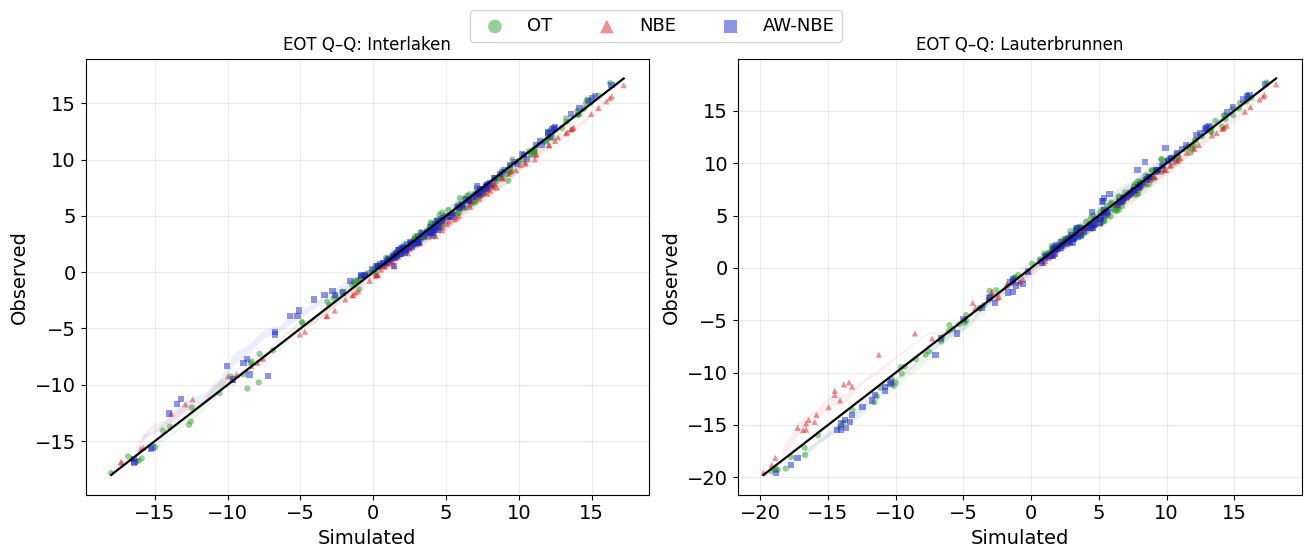}
\caption{\small \sf EOT Q--Q plots for the dry spell exceedances dataset under the MDGPD model.
The plots compare the observed samples with samples simulated from the fitted model under three estimation procedures the OT estimator in green circles, the NBE in red triangles and the AW--NBE in blue squares.
Each panel corresponds to the projection onto the $e_i$ coordinate, $i=1,2$, where $e_1$, $e_2$ correspond respectively to Interlaken and Lauterbrunnen.
The diagonal indicates perfect agreement between observed and simulated samples.}
\label{fig:QQplotDrySpellsComparison_OT_NBE_AW-NBE}
\end{figure}

Figure~\ref{fig:PotentialDrySpellsComparison} provides a diagnostic based on EOT potentials. The three estimators remain globally close to the diagonal for small and moderate values, showing that each fitted model captures the main transport structure of the data. Nevertheless, the NBE fit displays more visible departures from linearity as the exceedances increase, reflecting residual discrepancies between observed and simulated samples. 
Visually, the two transport-based methods (OT and AW--NBE) yield a similarly stable global alignment, clearly outperforming the baseline NBE. This highlights the relevance of the EOT fit method for retrieving the distribution, which will be further investigated through simulation analysis. Overall, the potential plots demonstrate that incorporating optimal transport into the fitting procedure results in a more accurate global transport alignment.

Figure~\ref{fig:QQplotDrySpellsComparison_OT_NBE_AW-NBE} additionally compares the multivariate EOT Q--Q plots obtained from the OT, NBE, and AW--NBE fits for the dry spell exceedances dataset. All three procedures reproduce the overall ordering structure of the exceedances reasonably well, as most points remain close to the diagonal over a large part of the support.
However, clearer differences emerge in the tails. The NBE estimator exhibits a systematic downward deviation in the upper quantiles, indicating an underestimation of the largest dry spell exceedances. The OT estimator provides a better overall alignment with the diagonal, especially for large values. By contrast, the AW--NBE fit is generally the closest to the diagonal across both stations and displays the smallest overall dispersion, particularly in the upper tail. This suggests that the Wasserstein refinement improves the agreement with the observed exceedances under the OT-based diagnostics.

To complement the graphical EOT diagnostics, we compare the NBE and AW--NBE fits applying the procedure in Section~\ref{sec:gof}. The results are presented in Table~\ref{tab:gof_dryspells}.
\begin{table}[H]
\centering
\caption{\small \sf Sinkhorn goodness-of-fit diagnostics for the dry spell exceedance dataset under the MDGPD model.}
\begin{tabular}{lcc}
\toprule
Method & $D_n(\widehat{\boldsymbol{\theta}}_n)$ & Bootstrap $p$-value \\
\midrule
NBE & 0.0596 & 0.56 \\
OT & 0.0324 & 0.31 \\
AW--NBE & 0.0186 & 0.42 \\
\bottomrule
\end{tabular}
\label{tab:gof_dryspells}
\end{table}
Table~\ref{tab:gof_dryspells} reports the Sinkhorn discrepancy $D_n(\widehat{\boldsymbol{\theta}}_n)$ and the associated bootstrap $p$-values. Although OT and AW--NBE show comparable performance in the visual potential plots, the numerical results reveal that the AW--NBE refinement achieves the best overall fit. Specifically, the AW--NBE estimator yields a discrepancy of $0.0186$, representing a nearly twofold improvement over OT ($0.0324$) and a threefold improvement over the baseline NBE ($0.0596$). 
All bootstrap $p$-values remain well above conventional significance levels, indicating that none of the fitted models is rejected by the goodness-of-fit test. 
Unlike the banking dataset, the dry spell data are discrete, and in this setting the OT estimator appears to outperform the NBE estimator. This suggests that discreteness may play an important role in the relative performance of the estimators and warrants further investigation. Moreover, comparing Tables \ref{tab:gof_banks} and \ref{tab:gof_dryspells} shows that the improvement is substantially more pronounced for the dry spell data than for the banking example.

\section{Conclusion}

This paper proposed a likelihood-free inference framework for MGPD models based on a two-step procedure combining neural Bayes estimation with an optimal transport refinement step. The AW--NBE estimator reduces the Sinkhorn discrepancy between observed and simulated exceedances while preserving the stability and computational efficiency of simulation-based neural inference.

Consistency was established under standard regularity conditions, and asymptotic normality under an additional local first-order expansion condition on the empirical Sinkhorn gradient. The verification of this condition for Sinkhorn-based simulation models remains an open theoretical question.

Empirical results on financial and environmental datasets show that the proposed method can improve the fit between observed and simulated exceedances compared with both censored likelihood estimation and baseline neural estimators, with gains that are particularly pronounced in the discrete setting. The proposed framework applies to both continuous and discrete MGPD models and provides a flexible alternative when likelihood evaluation is intractable.

Future work may extend the methodology to higher-dimensional settings, design adaptive tail-weighting strategies, and consider alternative discrepancies beyond Wasserstein-type ones. Additional directions include the theoretical analysis of Sinkhorn-based bootstrap calibration procedures and a deeper understanding of the finite-sample behavior observed in discrete models.

\paragraph{Acknowledgement.} 
The work of S. Aka was supported by ANRT (CIFRE PhD program) in collaboration with {\it Square Management}.
\\P. Naveau acknowledges financial support from the Agence Nationale de la Recherche (ANR) through the SICIM and SHARE PEPR Maths-Vives projects (France 2030, ANR-24-EXMA-0008), the EXSTA grant (ANR-23-CE40-0009-01), PORC-EPIC, the PEPR TRACCS program (PC4 EXTENDING, ANR-22-EXTR-0005), and the PEPR IRIMONT project (France 2030, ANR-22-EXIR-0003). He also benefited from the GeoLearning research chair.

\bibliographystyle{apalike}
\bibliography{Ref}

\newpage
\appendix

\section{Proofs}

\subsection{Proof of Theorem~\ref{thm:consistency_aw_nbe}}
\label{app:consistency_aw_nbe}

This proof relies on the standard separation argument for extremum estimators \cite[see, e.g.][]{NeweyMcFadden1994, vandervaart1998}. 

Let us define
\[
Q(\boldsymbol{\theta})
:=
S_\varepsilon(P_{\boldsymbol{\theta}_0},P_{\boldsymbol{\theta}}),
\qquad \boldsymbol{\theta}\in\Theta.
\]
For all $\boldsymbol{\theta}\in\Theta$, we can write
\[
Q_n(\boldsymbol{\theta})-Q(\boldsymbol{\theta})
=
\Bigl\{
S_\varepsilon(\widehat P_n,\widehat P_{m_n,\boldsymbol{\theta}})
-
S_\varepsilon(P_{\boldsymbol{\theta}_0},P_{\boldsymbol{\theta}})
\Bigr\}
+
\lambda_n\|\boldsymbol{\theta}-\widehat{\boldsymbol{\theta}}_n^{\mathrm{NBE}}\|_2^2,
\]
with $Q_n(\boldsymbol{\theta})$ introduced in Definition~\ref{def:AWNBE_sinkhorn}.\\
Hence,
\begin{equation}\label{eq:Q_n_inequality}
\sup_{\boldsymbol{\theta}\in\Theta}|Q_n(\boldsymbol{\theta})-Q(\boldsymbol{\theta})|
\le
\sup_{\boldsymbol{\theta}\in\Theta}
\Bigl|
S_\varepsilon(\widehat P_n,\widehat P_{m_n,\boldsymbol{\theta}})
-
S_\varepsilon(P_{\boldsymbol{\theta}_0},P_{\boldsymbol{\theta}})
\Bigr|
+
\lambda_n\sup_{\boldsymbol{\theta}\in\Theta}\|\boldsymbol{\theta}-\widehat{\boldsymbol{\theta}}_n^{\mathrm{NBE}}\|_2^2.    
\end{equation}
By compactness of $\Theta$, $R:=\sup_{\boldsymbol{\theta}\in\Theta}\|\boldsymbol{\theta}\|_2<\infty$.
Since we take $\widehat{\boldsymbol{\theta}}_n^{\mathrm{NBE}}\in\Theta$, then
$\|\widehat{\boldsymbol{\theta}}_n^{\mathrm{NBE}}\|_2\le R$.
Moreover, for any $\boldsymbol{\theta}\in\Theta$,
\[
\|\boldsymbol{\theta}-\widehat{\boldsymbol{\theta}}_n^{\mathrm{NBE}}\|_2
\le
\|\boldsymbol{\theta}\|_2+\|\widehat{\boldsymbol{\theta}}_n^{\mathrm{NBE}}\|_2
\le
2R,
\]
hence
\[
\sup_{\boldsymbol{\theta}\in\Theta}\|\boldsymbol{\theta}-\widehat{\boldsymbol{\theta}}_n^{\mathrm{NBE}}\|_2^2
\le (2R)^2 = 4R^2,
\]
and by (H4),
\[
\lambda_n\sup_{\boldsymbol{\theta}\in\Theta}\|\boldsymbol{\theta}-\widehat{\boldsymbol{\theta}}_n^{\mathrm{NBE}}\|_2^2
\le
4R^2\lambda_n
\to 0, \text{ as $n\to\infty$.}
\]
Combining this last result with (H3) in Equation~\eqref{eq:Q_n_inequality}, we obtain
\begin{equation}\label{eq:Q_n_sup_convergence}
\sup_{\boldsymbol{\theta}\in\Theta}|Q_n(\boldsymbol{\theta})-Q(\boldsymbol{\theta})|
\xrightarrow{\mathbb P}0 \text{, as $n\to\infty$.}
\end{equation}
Fix $\eta>0$ and define
\[
\Theta_\eta:=\{\boldsymbol{\theta}\in\Theta:\ \|\boldsymbol{\theta}-\boldsymbol{\theta}_0\|_2\ge\eta\},
\]
which is compact under (H1). 
By (H2), $Q(\boldsymbol{\theta})>0$ for every $\boldsymbol{\theta}\neq\boldsymbol{\theta}_0$,
so $\boldsymbol{\theta}_0$ is the unique minimizer of $Q$.
In particular, $Q(\boldsymbol{\theta})>0$ for all $\boldsymbol{\theta}\in\Theta_\eta$.
As $Q$ is continuous, 
\[
\gamma(\eta)
:=
\inf_{\boldsymbol{\theta}\in\Theta_\eta}Q(\boldsymbol{\theta})
=
\min_{\boldsymbol{\theta}\in\Theta_\eta}Q(\boldsymbol{\theta})
>0.
\]
Thus, for all $\boldsymbol{\theta}\in\Theta_\eta$,
\[
Q(\boldsymbol{\theta})\ge \gamma(\eta)
\quad \text{and} \quad
Q(\boldsymbol{\theta}_0)=0.
\]
Now, introduce the event
\[
A_n(\eta)
:=
\left\{
\sup_{\boldsymbol{\theta}\in\Theta}|Q_n(\boldsymbol{\theta})-Q(\boldsymbol{\theta})|
<
\frac{\gamma(\eta)}{3}
\right\},
\]
on which the uniform bound implies that for every $\boldsymbol{\theta}\in\Theta$,
\[
|Q_n(\boldsymbol{\theta})-Q(\boldsymbol{\theta})|
<
\frac{\gamma(\eta)}{3}. 
\]
Hence, for every $\boldsymbol{\theta}\in\Theta_\eta$,
\[
Q_n(\boldsymbol{\theta})
>
Q(\boldsymbol{\theta})-\frac{\gamma(\eta)}{3}
\ge
\gamma(\eta)-\frac{\gamma(\eta)}{3}
=
\frac{2\gamma(\eta)}{3}.
\]
Furthermore, at $\boldsymbol{\theta}_0$,
\[
Q_n(\boldsymbol{\theta}_0)
<
Q(\boldsymbol{\theta}_0)+\frac{\gamma(\eta)}{3}
=
\frac{\gamma(\eta)}{3}.
\]
Therefore, on $A_n(\eta)$,
\[
\inf_{\boldsymbol{\theta}\in\Theta_\eta}Q_n(\boldsymbol{\theta})
\ge
\frac{2\gamma(\eta)}{3}
>
\frac{\gamma(\eta)}{3}
\ge
Q_n(\boldsymbol{\theta}_0).
\]
Now, suppose by contradiction that $\widehat{\boldsymbol{\theta}}_n^{\mathrm{AW}}\in \Theta_{\eta}$. Then,
\[
Q_n(\widehat{\boldsymbol{\theta}}_n^{\mathrm{AW}})
\ge
\inf_{\boldsymbol{\theta}\in\Theta_\eta}Q_n(\boldsymbol{\theta})
>
Q_n(\boldsymbol{\theta}_0),
\]
which contradicts the minimizing property of $\widehat{\boldsymbol{\theta}}_n^{\mathrm{AW}}$.
Hence, on the event $A_n(\eta)$, one has
\[
\widehat{\boldsymbol{\theta}}_n^{\mathrm{AW}}\notin \Theta_{\eta}.
\]
This implies
\[
A_n(\eta)
\subseteq
\{\|\widehat{\boldsymbol{\theta}}_n^{\mathrm{AW}}-\boldsymbol{\theta}_0\|_2<\eta\}.
\]
By uniform convergence (Equation~\eqref{eq:Q_n_sup_convergence}),
\[
\mathbb P(A_n(\eta))\to 1
\quad\text{as } n\to\infty,
\]
which implies
\[
\widehat{\boldsymbol{\theta}}_n^{\mathrm{AW}}
\xrightarrow{\mathbb P}
\boldsymbol{\theta}_0.
\]

\subsection{Proof of Theorem~\ref{thm:asymptotics_AWNBE}}
\label{app:asymptotics_AWNBE}

We already know that as $n\to\infty$
$$
\widehat{\boldsymbol{\theta}}_n^{\mathrm{AW}} \xrightarrow{\mathbb P} \boldsymbol{\theta}_0 \text{ (Theorem~\ref{thm:consistency_aw_nbe})}\quad \text{and}
\quad \widehat{\boldsymbol{\theta}}_n^{\mathrm{EOT}} \xrightarrow{\mathbb P} \boldsymbol{\theta}_0 \text{ (\citealp{goldfeld2023statistical})}.
$$
Since $\widehat{\boldsymbol{\theta}}_n^{\mathrm{EOT}}$ is a local interior minimizer of
$\theta \mapsto S_\varepsilon(\widehat P_n,\widehat P_{m_n,\theta})$ which is, by $(H5)$, a differentiable function 
in $\mathcal U$, we have almost surely
$$
\nabla S_\varepsilon(\widehat P_n,\widehat P_{m_n,\widehat{\boldsymbol{\theta}}_n^{\mathrm{EOT}}})=0.
$$
Since $\widehat{\boldsymbol{\theta}}_n^{\mathrm{EOT}}$ is asymptotically normal at rate $\sqrt{n}$,
we have
$$
\widehat{\boldsymbol{\theta}}_n^{\mathrm{EOT}} - \boldsymbol{\theta}_0 = O_{\mathbb P}(n^{-1/2}).
$$
We now turn to $\widehat{\boldsymbol{\theta}}_n^{\mathrm{AW}}$.
Since $\widehat{\boldsymbol{\theta}}_n^{\mathrm{AW}}$ is a local minimizer of $Q_n$ (see Definition~ \ref{def:AWNBE_sinkhorn}), we have almost surely
$$
\nabla Q_n(\widehat{\boldsymbol{\theta}}_n^{\mathrm{AW}})=\nabla S_\varepsilon(\widehat P_n,\widehat P_{m_n,\widehat{\boldsymbol{\theta}}_n^{\mathrm{AW}}})
+
2\lambda_n
\bigl(
\widehat{\boldsymbol{\theta}}_n^{\mathrm{AW}}-\widehat{\boldsymbol{\theta}}_n^{\mathrm{NBE}}
\bigr)
=0.
$$
Applying the local expansion (H5) with $\boldsymbol{\theta}_n=\widehat{\boldsymbol{\theta}}_n^{\mathrm{AW}}$, we obtain
$$
0
=
\nabla S_\varepsilon(\widehat P_n,\widehat P_{m_n,\boldsymbol{\theta}_0})
+
\mathbf V_\varepsilon(\boldsymbol{\theta}_0)
\bigl(\widehat{\boldsymbol{\theta}}_n^{\mathrm{AW}}-\boldsymbol{\theta}_0\bigr)
+
2\lambda_n
\bigl(
\widehat{\boldsymbol{\theta}}_n^{\mathrm{AW}}-\widehat{\boldsymbol{\theta}}_n^{\mathrm{NBE}}
\bigr)
+
o_{\mathbb P}\!\left(
\|\widehat{\boldsymbol{\theta}}_n^{\mathrm{AW}}-\boldsymbol{\theta}_0\|_2 + n^{-1/2}
\right).
$$
Since both estimators take values in the compact set $\Theta$ (under $(H1)$), both
$\widehat{\boldsymbol{\theta}}_n^{\mathrm{AW}}$ and $\widehat{\boldsymbol{\theta}}_n^{\mathrm{NBE}}$
are bounded in probability, so
$$
\|\widehat{\boldsymbol{\theta}}_n^{\mathrm{AW}}-\widehat{\boldsymbol{\theta}}_n^{\mathrm{NBE}}\|_2=O_{\mathbb P}(1), 
$$
and, under $(H4^*)$,
\begin{equation}\label{eq:op_rac_half_AW_and_NBE}
2\lambda_n
\bigl(
\widehat{\boldsymbol{\theta}}_n^{\mathrm{AW}}-\widehat{\boldsymbol{\theta}}_n^{\mathrm{NBE}}
\bigr)
=
o_{\mathbb P}(n^{-1/2}).    
\end{equation}
Applying (H5) with $\boldsymbol{\theta}_n = \widehat{\boldsymbol{\theta}}_n^{\mathrm{EOT}}$, given its asymptotic properties (see \eqref{eq:EOT_asymptotics}), yields
$$
\nabla S_\varepsilon(\widehat P_n,\widehat P_{m_n,\boldsymbol{\theta}_0})
+
\mathbf V_\varepsilon(\boldsymbol{\theta}_0)
(\widehat{\boldsymbol{\theta}}_n^{\mathrm{EOT}}-\boldsymbol{\theta}_0)
=
o_{\mathbb P}(n^{-1/2}).
$$
Both expansions imply, after subtraction and using \eqref{eq:op_rac_half_AW_and_NBE},
\[
\mathbf V_\varepsilon(\boldsymbol{\theta}_0)
\bigl(
\widehat{\boldsymbol{\theta}}_n^{\mathrm{AW}}-\widehat{\boldsymbol{\theta}}_n^{\mathrm{EOT}}
\bigr)
=
o_{\mathbb P}\!\left(
\|\widehat{\boldsymbol{\theta}}_n^{\mathrm{AW}}-\boldsymbol{\theta}_0\|_2 + n^{-1/2}
\right).
\]
Using the triangle inequality and $\widehat{\boldsymbol{\theta}}_n^{\mathrm{EOT}}-\boldsymbol{\theta}_0=O_{\mathbb P}(n^{-1/2})$,
\[
\|\widehat{\boldsymbol{\theta}}_n^{\mathrm{AW}}-\boldsymbol{\theta}_0\|_2
\le
\|\widehat{\boldsymbol{\theta}}_n^{\mathrm{AW}}-\widehat{\boldsymbol{\theta}}_n^{\mathrm{EOT}}\|_2
+O_{\mathbb P}(n^{-1/2}),
\]
so that
\[
\mathbf V_\varepsilon(\boldsymbol{\theta}_0)
\bigl(
\widehat{\boldsymbol{\theta}}_n^{\mathrm{AW}}-\widehat{\boldsymbol{\theta}}_n^{\mathrm{EOT}}
\bigr)
=
o_{\mathbb P}\!\left(
\|\widehat{\boldsymbol{\theta}}_n^{\mathrm{AW}}-\widehat{\boldsymbol{\theta}}_n^{\mathrm{EOT}}\|_2
+n^{-1/2}
\right).
\]
Since $\mathbf V_\varepsilon(\boldsymbol{\theta}_0)$ is symmetric positive definite, let
$c:=\lambda_{\min}\bigl(\mathbf V_\varepsilon(\boldsymbol{\theta}_0)\bigr)>0$. Then
\[
c\,\|\widehat{\boldsymbol{\theta}}_n^{\mathrm{AW}}-\widehat{\boldsymbol{\theta}}_n^{\mathrm{EOT}}\|_2
\le
\bigl\|\mathbf V_\varepsilon(\boldsymbol{\theta}_0)
\bigl(\widehat{\boldsymbol{\theta}}_n^{\mathrm{AW}}-\widehat{\boldsymbol{\theta}}_n^{\mathrm{EOT}}\bigr)\bigr\|_2
=
o_{\mathbb P}\!\left(
\|\widehat{\boldsymbol{\theta}}_n^{\mathrm{AW}}-\widehat{\boldsymbol{\theta}}_n^{\mathrm{EOT}}\|_2
+n^{-1/2}
\right),
\]
which yields, after rearranging,
\[
\bigl(c-o_{\mathbb P}(1)\bigr)\,
\|\widehat{\boldsymbol{\theta}}_n^{\mathrm{AW}}-\widehat{\boldsymbol{\theta}}_n^{\mathrm{EOT}}\|_2
\le
o_{\mathbb P}(n^{-1/2}),
\]
and therefore
\[
\widehat{\boldsymbol{\theta}}_n^{\mathrm{AW}}-\widehat{\boldsymbol{\theta}}_n^{\mathrm{EOT}}
=
o_{\mathbb P}(n^{-1/2}).
\]
Then,
\[
\sqrt n(\widehat{\boldsymbol{\theta}}_n^{\mathrm{AW}}-\boldsymbol{\theta}_0)
=
\sqrt n(\widehat{\boldsymbol{\theta}}_n^{\mathrm{EOT}}-\boldsymbol{\theta}_0)
+
o_{\mathbb P}(1),
\]
from which we conclude that
\[
\sqrt n(\widehat{\boldsymbol{\theta}}_n^{\mathrm{AW}}-\boldsymbol{\theta}_0)
\xrightarrow{d}
\mathcal N\!\left(
0,
\mathbf V_\varepsilon(\boldsymbol{\theta}_0)^{-1}
\mathbf\Sigma_\varepsilon(\boldsymbol{\theta}_0)
\mathbf V_\varepsilon(\boldsymbol{\theta}_0)^{-1}
\right).
\]

\subsection{Proof of Proposition~\ref{prop:AW_vs_NBE}}
\label{app:AW_vs_NBE}

By definition of $\widehat{\boldsymbol{\theta}}_n^{\mathrm{AW}}$, we have
$$
\widehat{\boldsymbol{\theta}}_n^{\mathrm{AW}}
\in
\arg\min_{\boldsymbol{\theta}\in\Theta} Q_n(\boldsymbol{\theta}),
$$
which implies that for all $\boldsymbol{\theta}\in\Theta$,
$$
Q_n(\widehat{\boldsymbol{\theta}}_n^{\mathrm{AW}})
\le
Q_n(\boldsymbol{\theta}).
$$
Since $\widehat{\boldsymbol{\theta}}_n^{\mathrm{NBE}}\in\Theta$, we may choose $\boldsymbol{\theta}=\widehat{\boldsymbol{\theta}}_n^{\mathrm{NBE}}$, which yields
$$
Q_n(\widehat{\boldsymbol{\theta}}_n^{\mathrm{AW}})
\le
Q_n(\widehat{\boldsymbol{\theta}}_n^{\mathrm{NBE}}).
$$
But, by definition of $Q_n$, we have
$$
Q_n(\widehat{\boldsymbol{\theta}}_n^{\mathrm{NBE}})
=
S_\varepsilon(\widehat P_n,\widehat P_{m_n,\widehat{\boldsymbol{\theta}}_n^{\mathrm{NBE}}}) \quad \text{and}\quad Q_n(\widehat{\boldsymbol{\theta}}_n^{\mathrm{AW}})
=
S_\varepsilon(\widehat P_n,\widehat P_{m_n,\widehat{\boldsymbol{\theta}}_n^{\mathrm{AW}}})
+
\lambda_n
\|\widehat{\boldsymbol{\theta}}_n^{\mathrm{AW}}-\widehat{\boldsymbol{\theta}}_n^{\mathrm{NBE}}\|_2^2.
$$
Hence, we obtain
$$
S_\varepsilon(\widehat P_n,\widehat P_{m_n,\widehat{\boldsymbol{\theta}}_n^{\mathrm{AW}}})
+
\lambda_n
\|\widehat{\boldsymbol{\theta}}_n^{\mathrm{AW}}-\widehat{\boldsymbol{\theta}}_n^{\mathrm{NBE}}\|_2^2
\le
S_\varepsilon(\widehat P_n,\widehat P_{m_n,\widehat{\boldsymbol{\theta}}_n^{\mathrm{NBE}}}),
$$
from which, we deduce that
$$
S_\varepsilon(\widehat P_n,\widehat P_{m_n,\widehat{\boldsymbol{\theta}}_n^{\mathrm{AW}}})
\le
S_\varepsilon(\widehat P_n,\widehat P_{m_n,\widehat{\boldsymbol{\theta}}_n^{\mathrm{NBE}}}).
$$

\section{Multivariate GPD}
\label{sec:mgpds_def}

This section recalls the notion of multivariate generalized Pareto distributions in both continuous and discrete settings. We first review the representations and properties of continuous MGPDs, which arise as limiting laws of multivariate threshold exceedances. We then introduce their discrete counterparts, the MDGPDs, designed to model integer-valued extremes while preserving threshold stability and dependence. These constructions provide the toy models for the applications in Section~\ref{sec:real_data}.

\subsection{Continuous MGPD}

Multivariate generalized Pareto distributions (MGPDs) arise as limiting distributions for multivariate threshold exceedances and form a central object of multivariate extreme value theory. They extend the univariate generalized Pareto distribution to the multivariate setting and are closely connected to multivariate extreme value (max-stable) distributions; see \citet{rootzen_tajvidi2006,rootzen_segers_wadsworth,rootzenSegerswadsworth_2}.

Let $\mathbf{X}=(X_1,\dots,X_d)$ be a random vector with distribution function $F$ defined on $\mathbb{R}^d$. For a threshold vector $\mathbf{u}\in\mathbb{R}^d$, an observation $\mathbf{x}$ is said to exceed $\mathbf{u}$ if at least one component exceeds its marginal threshold, that is,
$$
\mathbf{x} \nleq \mathbf{u},
$$
where inequalities between vectors are interpreted componentwise. The corresponding exceedance region
$$
\{\mathbf{x}\in\mathbb{R}^d : \mathbf{x}\nleq \mathbf{u}\}
$$
has an L-shaped geometry, reflecting that extreme events may occur through the exceedance of any marginal component.

If $F$ belongs to the multivariate domain of attraction of a multivariate extreme value distribution $G$, then, under suitable marginal normalizations, the conditional distribution of threshold exceedances converges to a multivariate generalized Pareto distribution. Specifically, letting
$$
\mathbf{X}_{\mathbf{u}}
=
\frac{\mathbf{X}-\mathbf{u}}
{\boldsymbol{\sigma}(\mathbf{u})},
$$
we have
$$
\pi\!\left(
\mathbf{X}_{\mathbf{u}}
\mid
\mathbf{X}_{\mathbf{u}}\nleq \mathbf{0}
\right)
\Longrightarrow H,
$$
where $H$ is a MGPD associated with $G$ \citep{rootzen_tajvidi2006}.

Following \citet{rootzen_segers_wadsworth,rootzenSegerswadsworth_2}, a convenient characterization of MGPDs is obtained through marginal standardization. A random vector $\mathbf{Z}=(Z_1,\dots,Z_d)$ defined on $[-\infty,\infty)^d$ follows a standard MGPD if it admits the representation
$$
\mathbf{Z} =\mathbf{S}+ E,
$$
where $E$ is a unit exponential random variable and $\mathbf{S}$ is a random vector taking values in $[-\infty,0]^d$, independent of $E$, and satisfying $\max(\mathbf{S})=0$ almost surely.

The support of $\mathbf{Z}$ is contained in $
\{\mathbf{z}\in[-\infty,\infty)^d : \max(\mathbf{z})>0\}
$. This representation separates the magnitude of exceedances, governed by the exponential variable $E$, from the extremal dependence structure encoded by the spectral vector $\mathbf{S}$ \cite[see, e.g., Chapter~7]{HandbookExtremes2026}.

The spectral vector can be constructed by introducing an auxiliary random vector $\mathbf{T}$ on $[-\infty,\infty)^d$ such that $\max(\mathbf{T})>- \infty$ almost surely and defining

\begin{equation}\label{eq:T_generator}
\mathbf{S}=\mathbf{T}-\max(\mathbf{T}).    
\end{equation}
Different choices of the distribution of $\mathbf{T}$ lead to different dependence structures.

A general MGPD is obtained through marginal transformation of a standard MGPD. A random vector $\mathbf{X}$ follows a MGPD with scale vector $\boldsymbol{\sigma}\in(0,\infty)^d$ and shape vector $\boldsymbol{\xi}\in\mathbb{R}^d$ if
$$
\mathbf{X}
=
\boldsymbol{\sigma}
\frac{\exp(\boldsymbol{\xi}\mathbf{Z})-\mathbf{1}}
{\boldsymbol{\xi}},
$$
where operations are interpreted componentwise and the usual limit convention is used when $\xi_j=0$.

The support of each marginal component depends on the sign of $\xi_j$. 
If $\xi_j>0$, then $X_j$ is bounded below by $-\sigma_j/\xi_j$. 
If $\xi_j=0$, the support is unbounded. 
If $\xi_j<0$, then $X_j$ is bounded above by $-\sigma_j/\xi_j$.

MGPD distributions satisfy a threshold stability property: if $\mathbf{X}$ follows a MGPD and $\mathbf{v}\ge \mathbf{0}$ is such that $\boldsymbol{\sigma}+\boldsymbol{\xi}\mathbf{v}>0$ componentwise, then
$$
(\mathbf{X}-\mathbf{v}\mid \mathbf{X}\nleq \mathbf{v})
$$
is again MGPD with updated scale parameters and identical shape parameters \citep{rootzen_segers_wadsworth}.

\subsection{Discrete MGPD}

While MGPDs provide a powerful framework for modeling multivariate extremes in the continuous setting, discrete distributions generally do not possess a classical maximum domain of attraction, which complicates the extension of multivariate extreme value theory to integer-valued settings. Recent developments have therefore aimed to construct discrete analogues of extreme value models that retain their key asymptotic and stability properties. In this spirit, multivariate discrete generalized Pareto distributions (MDGPDs) were introduced in \cite{aka2025multivariatediscretegeneralizedpareto} as an extension of MGPDs to integer-valued data. Unlike discretizations of continuous models, MDGPDs are defined directly on $\mathbb{Z}^d$ and preserve the fundamental structural properties of multivariate threshold exceedances.

Following \citet{aka2025multivariatediscretegeneralizedpareto}, a standard MDGPD random vector $\mathbf{N}$ is constructed through a stochastic representation analogous to the continuous spectral decomposition of MGPDs. In this setting, the maximum component
$$
G = \max(\mathbf{N})
$$
follows a geometric distribution, which plays the discrete counterpart of the exponential radial component appearing in continuous MGPD representations. The associated discrete spectral vector
$$
\mathbf{S} = \mathbf{N} - G
$$
is independent of $G$ and satisfies $\max(\mathbf{S}) = 0$. This decomposition separates marginal tail from extremal dependence, mirroring the continuous MGPD structure.

This leads to the representation
$$
\mathbf{N} = \mathbf{T} - \max(\mathbf{T}) + G,
$$
where $\mathbf{T}$ is a discrete generator encoding the dependence structure of the model. The generator $\mathbf{T}$ plays a central role in shaping extremal dependence, allowing for a wide range of dependence patterns through flexible parametric or nonparametric specifications. In particular, the distribution of pairwise differences induced by $\mathbf{T}$ governs the joint extremal behavior of the MDGPD.

An important consequence of this construction is that MDGPDs preserve a discrete analogue of the threshold stability property. Specifically, the marginal conditional exceedance distribution satisfies a geometric memoryless property, ensuring that exceedances above increasing thresholds remain within the same parametric family. This property provides a discrete counterpart to the classical POT characterization of MGPDs.

Non-standard MDGPDs are obtained by defining their cumulative distribution function through a discrete transformation of the continuous MGPD representation. Since the marginal transformation used in the continuous setting does not preserve the integer support, it cannot be directly applied in the discrete case.

We therefore define non-standard MDGPDs through their cumulative distribution function. A random vector $\mathbf{M} \in \mathbb{Z}^d$ is said to follow a non-standard MDGPD with parameters $(\boldsymbol{\sigma}, \boldsymbol{\xi}, \mathbf{S})$ if, for all $\mathbf{k} \in \mathbb{Z}^d$,
\begin{equation*}
\mathbb{P}(\mathbf{M} \leq \mathbf{k})
=
1 - \mathbb{E}
\left[
1 \wedge \exp\left(
\max\left(
\mathbf{S}
-
\frac{1}{\boldsymbol{\xi}}
\log\bigl(1 + \boldsymbol{\xi}\,\mathbf{k}/\boldsymbol{\sigma}\bigr)
\right)
\right)
\right].  
\end{equation*}
This construction yields a flexible family indexed by marginal scale and shape parameters $(\boldsymbol{\sigma}, \boldsymbol{\xi})$, together with a discrete spectral distribution governing extremal dependence.

Importantly, MDGPDs are not mere discretizations of MGPDs. Although asymptotic equivalence holds when the scale parameter tends to infinity, finite-sample behavior can differ substantially, particularly when tail probabilities are influenced by integer constraints and lattice effects. This justifies the need for a dedicated discrete modeling framework.

From an inferential perspective, MDGPDs pose significant challenges because their likelihood is generally intractable, especially when the generator distribution is unknown or high-dimensional. This naturally motivates the use of likelihood-free inference methods, such as NBEs or optimal transport–based approaches, which rely on faithful simulation of the underlying discrete model.

\section{EOT Q--Q and potential plots for comparison between likelihood-free inference strategies (banks log-returns and dry spells)}\label{appendix:simulation}

\subsection{Banks log-returns from \texorpdfstring{\cite{Kiriliouk2019}}{Kiriliouk et al. (2019)}}

\begin{figure}[H]
\centering
\includegraphics[width=0.5\linewidth]{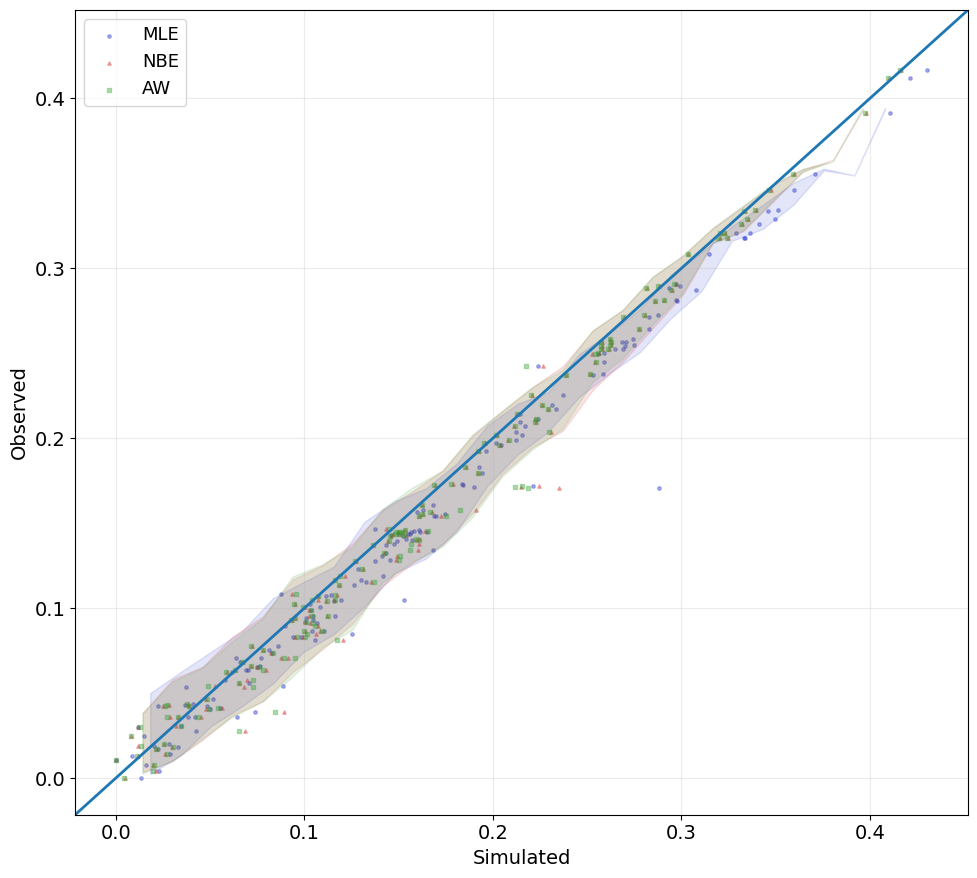}
\caption{\small \sf EOT potential plots with 95\% bootstrap confidence bands for the banking return exceedances dataset under three estimation methods MLE in blue, NBE in red, AW--NBE in green for the Gumbel--$\boldsymbol{T}$ MGPD model. The potentials are centered to remove arbitrary additive constants. All methods are computed using a common entropic regularization parameter $\varepsilon = 10^{-2}$.}
\label{fig:PotentialBanksComparison_CI}
\end{figure}

\begin{figure}[H]
\centering
\hspace{-5ex}
\includegraphics[width=1.05\linewidth]{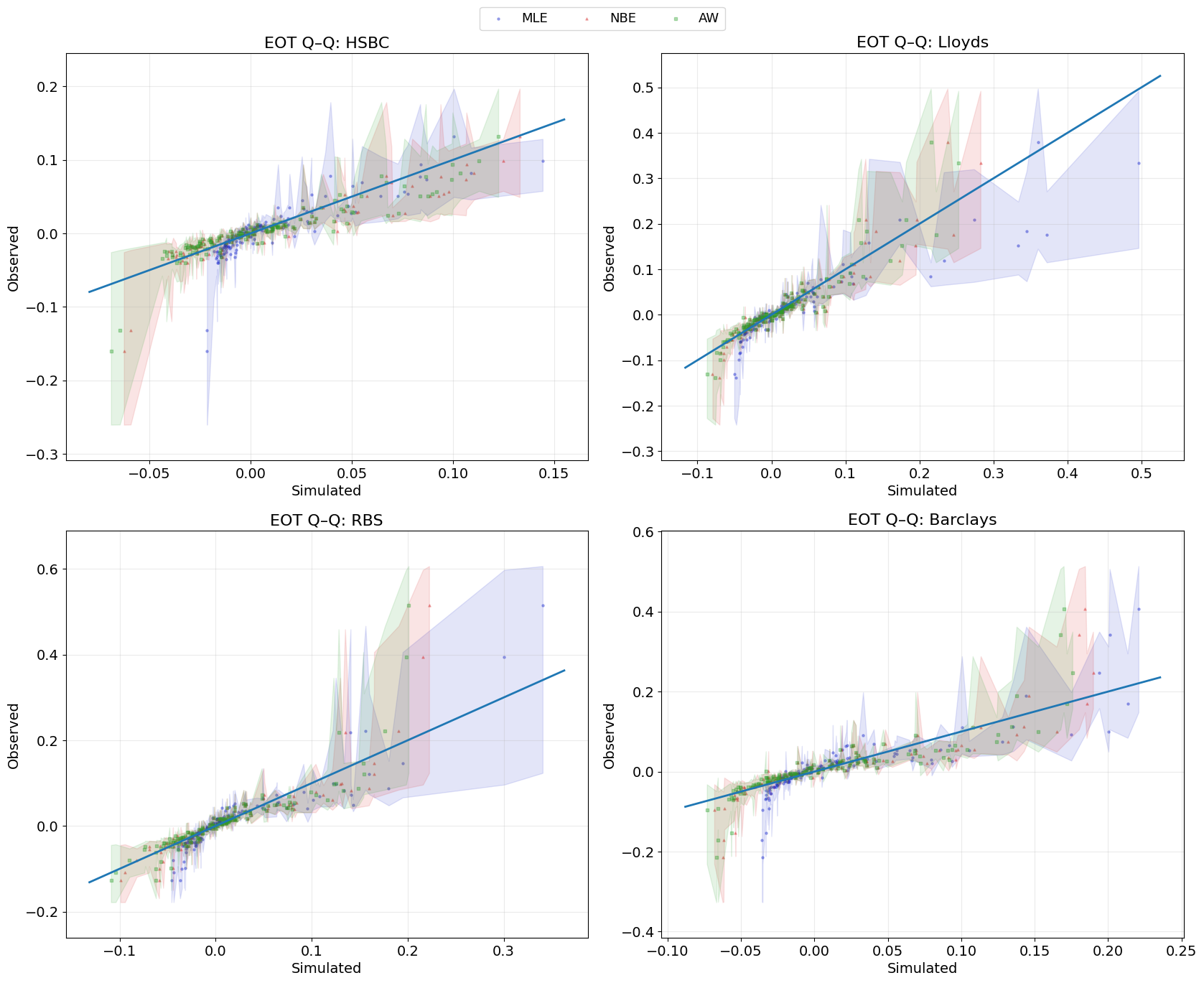}
\caption{\small\sf Quadrivariate EOT Q--Q plots with 95\% bootstrap confidence bands for the banking return exceedances dataset under three estimation methods MLE in blue, NBE in red, AW--NBE in green for the Gumbel--$\boldsymbol{T}$ MGPD model. Each panel corresponds to the projection onto the $e_i$ coordinate, $i=1,\ldots,4$, where $e_1$, $e_2$, $e_3$, and $e_4$ correspond respectively to HSBC, Lloyds, RBS, and Barclays. EOT quantiles are computed using a shared entropic regularization parameter $\varepsilon = 10^{-2}$ to ensure fair visual comparison across methods.}
\label{fig:QQplotBanksComparison_CI}
\end{figure}

\begin{figure}[H]
\centering
\includegraphics[width=0.5\linewidth]{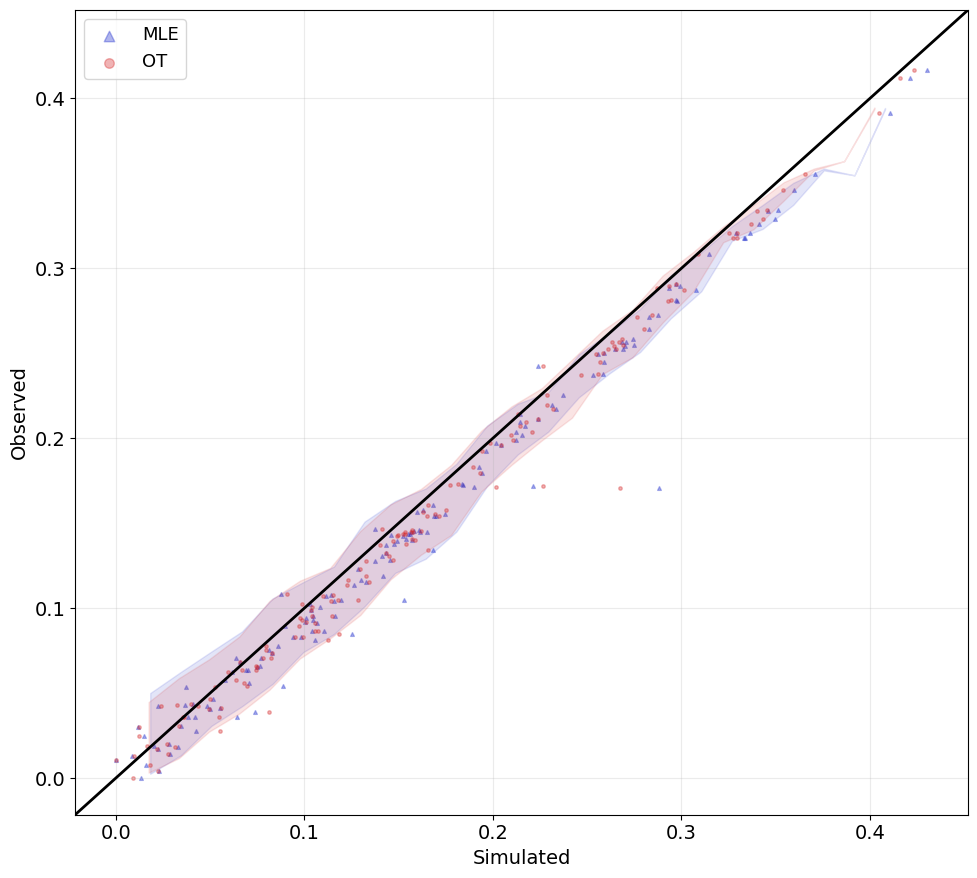}
\caption{\small\sf
EOT potential plots for the banks log-returns exceedances dataset under the MGPD model.
The plots compare the MLE in blue triangles and the OT estimator in red circles.}
\label{fig:PotentialBanksMLEOT}
\end{figure}

\begin{figure}[H]
\centering
\includegraphics[width=0.95\linewidth]{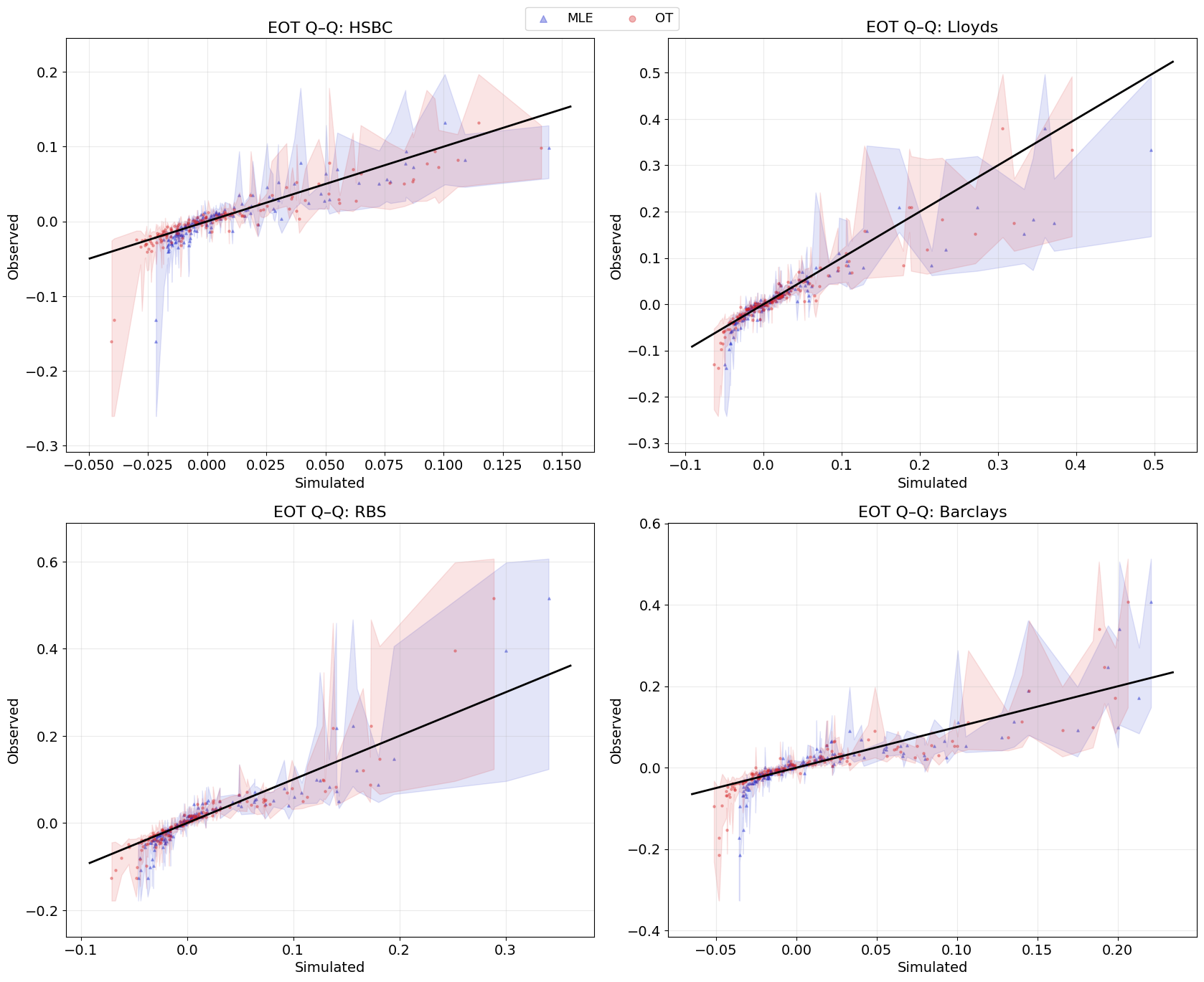}
\caption{\small\sf
EOT Q--Q plots for the bank log-return exceedances dataset under the MGPD model.
The plots compare observed samples with samples simulated from the fitted model using two estimation procedures MLE in blue triangles and the OT estimator in red circles.
Each panel corresponds to the projection onto the $e_i$ coordinate, $i=1,\ldots,4$, where $e_1$, $e_2$, $e_3$, and $e_4$ correspond respectively to HSBC, Lloyds, RBS, and Barclays.
}
\label{fig:QQplotBanksMLEOT}
\end{figure}

This section provides the full set of EOT graphical diagnostics 
for the banking dataset. Figures~\ref{fig:PotentialBanksComparison_CI} and~\ref{fig:QQplotBanksComparison_CI} 
compare MLE, NBE, and AW--NBE via EOT potential and Q--Q plots, 
while Figures~\ref{fig:PotentialBanksMLEOT} and~\ref{fig:QQplotBanksMLEOT} focus on the pairwise 
comparison between MLE and the OT estimator.\\[1ex]
Figure~\ref{fig:PotentialBanksComparison_CI} provides complementary diagnostics based on the potentials associated with the entropic optimal transport coupling between observed and simulated excess samples. The potentials summarize global transport discrepancies between the empirical and simulated distributions and therefore provide an alternative goodness-of-fit diagnostic. For the MLE estimator, the potentials follow an approximately linear pattern but exhibit visible departures from the diagonal at intermediate and large quantile levels, indicating residual discrepancies in the joint tail representation. The NBE estimator substantially reduces these discrepancies. The potentials concentrate more tightly around the diagonal, although moderate deviations remain in the lower tail region. These may reflect the finite-threshold nature of the peaks-over-threshold approximation and possible second-order tail effects, particularly given the limited number of joint exceedances. The AW--NBE estimator further improves the fit. The potentials remain tightly concentrated around the diagonal across most quantile levels, indicating reduced global transport discrepancies between observed and simulated excess distributions. Minor deviations persist in the lower tail region and are consistent with finite-sample variability.
Figure~\ref{fig:QQplotBanksComparison_CI} evaluates the adequacy of the three estimators by comparing observed excesses with excesses simulated from the fitted Gumbel--$\boldsymbol{T}$ MGPD model. Both samples contain the same number of joint exceedances ($n=149$) and are equally weighted. The panels display EOT Q--Q plots together with pointwise 95\% bootstrap confidence bands.

Overall, all estimators capture the global tail behavior satisfactorily. However, systematic differences appear across methods. The MLE fit exhibits noticeable curvature in several margins, particularly for the largest quantiles, suggesting residual discrepancies between observed and simulated extremes and indicating that the parametric likelihood-based specification may not fully reproduce the extremal dependence structure. The NBE estimator reduces these deviations. The Q--Q plots show improved alignment with the diagonal across most quantile levels, together with narrower confidence bands. This indicates that the likelihood-free estimator provides a more flexible representation of the extremal dependence structure while preserving adequate marginal tail behavior. The AW--NBE estimator in green provides the closest agreement with the diagonal, particularly in the upper quantile region. The dispersion of the EOT quantile pairs around the diagonal is further reduced, suggesting improved reproduction of joint exceedance patterns and a more accurate description of extremal dependence.\\[1ex]
Overall, the graphical diagnostics consistently suggest a progressive improvement from MLE to NBE and moderate one from NBE to AW--NBE. 

Figure~\ref{fig:PotentialBanksMLEOT} shows that, especially in the tail, the MLE exhibits stronger deviations, while the OT estimator remains closer to the diagonal.

When turning to Figure~\ref{fig:QQplotBanksMLEOT}, the difference in behavior between the OT estimator and the MLE is less pronounced than in Figure~\ref{fig:PotentialBanksMLEOT}, although the points and confidence region associated with the MLE appear shifted to the right along the diagonal.
~\hspace{-5ex}

\subsection{Dry spells from \cite{aka2025multivariatediscretegeneralizedpareto}}

\begin{figure}[H]
\centering
\includegraphics[width=0.7\linewidth]{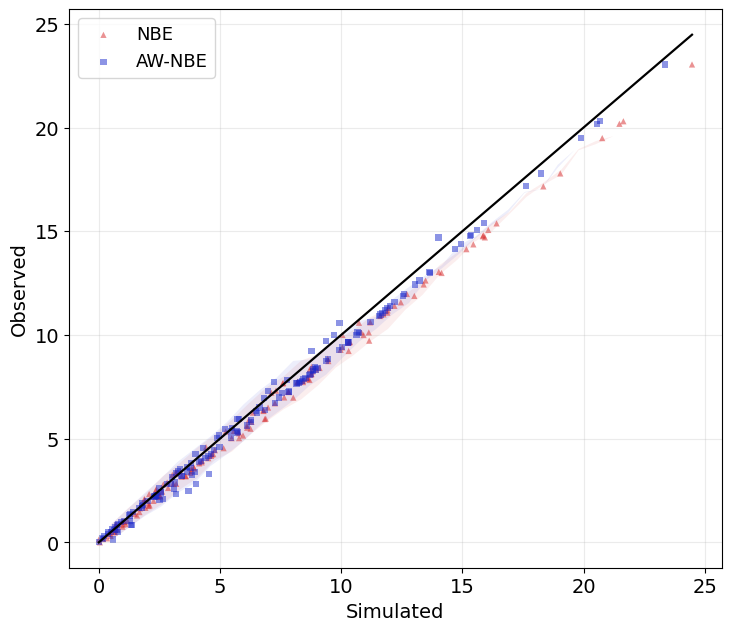}
\caption{\small\sf
EOT potentials for the dry spell exceedances dataset under the MDGPD model.
The red triangles correspond to the NBE estimator and the blue squares to the AW--NBE estimator.
}
\label{fig:PotentialDrySpellsComparison_NBE_AW-NBE}
\end{figure}
\begin{figure}[H]
\centering
\hspace{-5ex}
\includegraphics[width=1.05\linewidth]{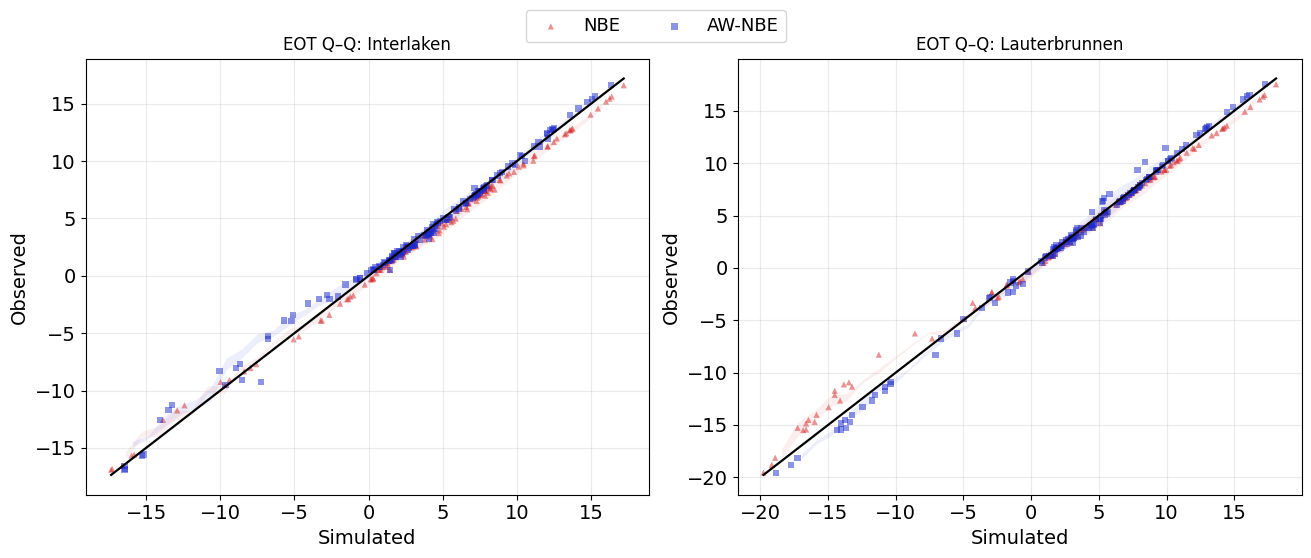}
\caption{\small\sf
Bivariate EOT Q--Q plots for the dry spell exceedances dataset under the MDGPD model.
The plots compare observed samples with samples simulated from the fitted model using the optimal transport coupling.
Red triangles correspond to the NBE estimator and blue squares to the AW--NBE estimator.
Each panel corresponds to the projection onto the $e_i$ coordinate, $i=1,2$, where $e_1$, $e_2$ correspond respectively to Interlaken and Lauterbrunnen.
The diagonal indicates perfect agreement between observed and simulated samples.
}
\label{fig:QQplotDrySpellsComparison_NBE_AW-NBE}
\end{figure}

In Figure~\ref{fig:PotentialDrySpellsComparison_NBE_AW-NBE}, we observe that the AW--NBE potentials points are more tightly concentrated around the diagonal, indicating a more stable transport coupling between observed and simulated exceedances. By contrast, the NBE potentials display larger curvature and visible departures from linearity for intermediate and large exceedances, reflecting residual discrepancies in the transport structure. Overall, the potential plots confirm the improved global alignment achieved by the AW--NBE refinement.\\[1ex]
Complementary, Figure~\ref{fig:QQplotDrySpellsComparison_NBE_AW-NBE} compares the multivariate EOT quantiles obtained from the NBE and AW--NBE fits for the dry spell exceedances dataset. Both estimators reproduce the global ordering structure of the exceedances. For moderate quantile levels, the two methods are close to the diagonal, even though NBE is in better agreement than AW--NBE in the lower tail. However, the NBE estimator exhibits a systematic downward deviation in the upper quantiles, indicating an underestimation of the largest dry spell exceedances. In contrast, the AW--NBE fit aligns more closely with the diagonal across most quantile levels and shows reduced dispersion, particularly in the upper tail, which is of greater interest to us.

\end{document}